\documentclass[onecolumn,superscriptaddress,
 amsmath,amssymb,
 aps,
 prd,
]{revtex4}
\usepackage{dcolumn}
\usepackage{bm}
\usepackage{graphicx}
\usepackage{physics}
\usepackage{tikz}
\usepackage{mathtools}
\usepackage{hyperref}%
\usepackage{comment}
\usepackage[caption=false]{subfig}
\usepackage{color}

\graphicspath{{FIGURES}}

\usepackage[ruled,lined]{algorithm2e}
\usepackage{setspace}
\definecolor{crimson}{HTML}{DC143C}
\definecolor{perfect_green}{HTML}{4FBF26}

\newcommand{\bea}{\begin{eqnarray}}
\newcommand{\eea}{\end{eqnarray}}

\begin{document}

\title{Resetting by rescaling: exact results for a diffusing particle in one-dimension}
\author{Marco Biroli}
\affiliation{LPTMS, CNRS, Univ.  Paris-Sud,  Universit\'e Paris-Saclay,  91405 Orsay,  France}
\author{Yannick Feld}
\affiliation{Universit\'e Paris-Saclay, CNRS, CEA, Institut de Physique Th\'eorique, 91191 Gif-sur-Yvette, France}
\author{Alexander K. Hartmann}
\affiliation{Institut für Physik, Carl von Ossietzky Universität Oldenburg, 26111 Oldenburg, Germany}
\author{Satya N. Majumdar}
\affiliation{LPTMS, CNRS, Univ.  Paris-Sud,  Universit\'e Paris-Saclay,  91405 Orsay,  France}
\author{Gr\'egory Schehr}
\affiliation{Sorbonne Universit\'e, Laboratoire de Physique Th\'eorique et Hautes Energies, CNRS UMR 7589, 4 Place Jussieu, 75252 Paris Cedex 05, France}


\begin{abstract}
In this paper, we study a simple model of a diffusive particle on a line, undergoing a stochastic resetting with rate $r$, 
via rescaling its current position by a factor $a$, which can be either positive or negative. For $|a|<1$, the position distribution
becomes stationary at long times and we compute this limiting distribution exactly for all $|a|<1$. This symmetric distribution 
has a Gaussian shape near its peak at $x=0$, but decays exponentially for large $|x|$. We also studied  
the mean first-passage time (MFPT) $T(0)$ to a target located at a distance $L$ from the initial position (the origin) of the particle. 
As a function of the initial position $x$, the MFPT $T(x)$ satisfies a nonlocal second order differential equation and we have solved
it explicitly for $0 \leq a < 1$. For $-1<a\leq 0$, we also solved it analytically but up to a constant factor $\kappa$ whose value can
be determined independently from numerical simulations. Our results show that, for all $-1<a<1$, the MFPT $T(0)$ (starting from the origin) 
shows a minimum at $r=r^*(a)$. However, the optimised MFPT $T_{\rm opt}(a)$ turns out to be a monotonically increasing function of $a$
for $-1<a<1$. This demonstrates that, compared to the standard resetting to the
origin ($a=0$), while the positive rescaling is not beneficial for the search of a target, the negative rescaling is. Thus resetting via rescaling followed by a reflection around the origin expedites the search of a target in one dimension.  
\end{abstract}

\maketitle

\section{Introduction} \label{sec:introduction}

Search processes are ubiquitous in nature, e.g., animals searching for food, a helicopter searching for survivors after a ship wreck, a protein
searching for a site to bind on a DNA strand, or even finding a bug in a computer program~\cite{VLRS11,BLMV11}. In many of these search processes, natural
intuition tells us that, if one is unsuccessful in finding a target for a while, perhaps one should stop the search and restart from the beginning. The
rationale behind this intuition is that, if one restarts the search, perhaps one will find a new pathway that leads to the target in a shorter time. 
This idea was given a concrete form by studying a simple analytical toy model of a particle diffusing and subjected to a stochastic resetting to its initial position with a constant rate $r$. For example, in one dimension, let us consider a particle with position $x(t)$ at time $t$ that starts from $x(0) = 0$ and the position gets updated
by the rule \cite{EM11,EM11b}
\bea \label{def_RBM}
x(t+\dd t) = 
\begin{cases}
&x(0) =  0 \quad, \quad \hspace*{1.55cm}{\rm with \; probability} \quad r\, \dd t  \\
&x(t) + \sqrt{2 D\, \dd t} \, \xi(t) \, \quad, \quad {\rm with \; probability} \quad 1- r\, \dd t \;,
\end{cases}
\eea 
where $D$ is the diffusion constant and $\xi(t)$, for each $t$, are independent Gaussian random variables of zero mean and unit variance. In the limit $\dd t \to 0$, the increment $\xi(t)$ can be replaced by $\sqrt{\dd t}\,\eta(t)$,
resulting in the stochastic differential equation $dx/dt=\sqrt{2D}\eta(t)$.
Here $\eta(t)$ is the standard Gaussian white noise with zero mean and a delta-correlator $\langle \eta(t) \eta(t') \rangle = \delta(t-t')$. This simple toy model leads to two principle paradigms. 

The first natural question is: what is the position distribution $P_{r}(x,t)$ at time $t$? When $r=0$, $P_{r=0}(x,t)= e^{-x^2/(4 Dt)}/\sqrt{4 \pi D\,t}$ is simply a Gaussian. One of the first effects of resetting to the origin is that it breaks the detailed balance and drives the system to a nonequilibrium steady state (NESS) where the position distribution becomes non-Gaussian and is given by \cite{EM11,EM11b}
\bea \label{P_RBM}
P_r(x) = \lim_{t \to \infty} P_r(x,t) = \frac{1}{2} \sqrt{\frac{r}{D}}\, e^{- \sqrt{\frac{r}{D}}\,|x|} \;.
\eea
This theoretical prediction has been verified experimentally, using holographic optical tweezers \cite{TPSRR20}. The emergence of a new type of NESS induced by resetting is the first paradigm.

The second natural question is whether such a resetting really helps in reducing the search time of a target. Indeed, if one puts a target at a fixed distance $L$ from the origin, the mean first-passage time (MFPT) to the target (a simple measure of the search time) can be computed exactly. For a particle starting at and resetting to the origin, this MFPT is given by \cite{EM11,EM11b}
\bea \label{MFPT_RBM}
T(0) =  \frac{1}{r} \left[e^{\sqrt{\frac{r}{D}}\, L} - 1\right] \;,
\eea
where the argument $'0'$ refers to the starting position, i.e., the origin $x=0$. In the absence of resetting, the MFPT is infinite, while it becomes finite for any nonzero resetting rate $r$. Moreover, as a function of $r$, the MFPT exhibits a minimum at $r=r^*$, indicating the existence of an optimal resetting rate. Indeed, from Eq. (\ref{MFPT_RBM}), the dimensionless MFPT defined as 
\bea \label{def_Tilde}
\tilde T(0) = \frac{D T(0)}{L^2} \;,
\eea
can be expressed as a function of the dimensionless parameter $\beta = \sqrt{r/D} L$ as
\bea \label{expl_Tilde}
\tilde T(0) = \frac{1}{\beta^2} \left(e^{\beta} - 1 \right) \;.
\eea
As a function of $\beta$, it has a minimum at $\beta = 1.59362\cdots$. This clearly shows that in this simple toy model, switching on the resetting is not only advantageous to find a target, but there even exists an optimal resetting rate
that optimises the search time. This is the second paradigm of this toy model. 

These two paradigms, deduced from this simple toy model, 
have subsequently been found in several stochastic processes with resetting. The conditions for the existence of a NESS as well as an optimal $r^*$
have been studied in a variety of situations. Furthermore, some of these analytical predictions have been verified in experiments using optical trap
setups. For recent reviews of these developments, see \cite{EMS20,GJ2022,PKR2022} .

In this original toy model, the walker's position is always reset to the origin. However, this model can be analytically solved even when the reset position is not fixed after every reset, but rather is drawn independently from a distribution after each reset \cite{EM11b}. In fact, in the experimental setups used in Refs. \cite{BBPMC20,FBPCM21}, the resetting position typically corresponds to the Boltzmann distribution of the particle confined in a potential $U(x)$. However, the two paradigms discussed above, namely the existence of a NESS and a finite MFPT remain true when the resetting position is drawn from a distribution. In these models, the position after the resetting is uncorrelated with the position before reset, i.e., the post-resetting position has no memory of the pre-resetting position. A simple model that retains some dependence on the pre-resetting position has recently been studied under the name of  ``backtrack resetting'' where the evolution rule (\ref{def_RBM}) is modified to~\cite{TRR22,P22}
\begin{equation}\label{eq:summary-model-dynamics}
x(t + \dd t) = \begin{cases}
a \, x(t) &\mbox{~~with probability~~} r \dd t \;, \\
x(t) + \sqrt{2 D\, \dd t} \, \xi(t)\, & \mbox{~~with probability~~} 1 - r \dd t \;,
\end{cases}
\end{equation}
where $0 \leq a \leq 1$ is a backtracking parameter. This model represents for example the situation where $x(t)$ denotes the population of a habitat at time $t$, a fraction of which gets wiped out after a catastrophic event that occurs at random times distributed via a Poisson distribution with rate $r$ \cite{TRR22}. This model for $a>0$ sometimes goes by the name of ``partial resetting'' \cite{TRR22,P22,BCHPM23,OG24}. For $a=0$, this model reduces to the standard model of diffusion with stochastic resetting defined in Eq. (\ref{def_RBM}). For $a=1$, clearly the post and pre-resetting positions are the same (thus effectively there is no resetting) and the particle simply diffuses. 
For $a > 1$ the particle eventually escapes to $\pm \infty$. Thus there is no stationary state for $a \geq 1$ and the NESS exists only for $0\leq a<1$. 
In this range of $a$, the stationary position distribution has been computed in several recent papers \cite{TRR22,P22,BCHPM23,OG24,H23}. However, to the best of our knowledge, no result for the MFPT exists for general $a > 0$. In all these studies on the ``partial resetting'' model above, the parameter $a$ was considered to be nonnegative $a \geq 0$. However, in principle, one can study resetting for $a<0$ also \cite{H23}. A negative value of $a$ means a partial resetting coupled with a reflection around the origin. It is clear that for $a \leq -1$, there is no stationary state and the stationary state exists only in the range $-1<a \leq 0$ \cite{H23}. However, the stationary position distribution for $-1<a<0$ is not known explicitly. Furthermore, the MFPT has also not been studied for $-1<a<0$. To summarize, for this partial resetting model with parameters $-1<a<+1$, the position distribution does become stationary at late times and is known explicitly only in the range $0 \leq a < 1$. The MFPT is essentially unknown in the full range $-1< a < + 1$, except for $a=0$. One interesting open question is whether this additional parameter $a$ can reduce the MFPT compared the standard $a=0$ model of stochastic resetting. 

In this paper, we revisit this problem and compute analytically the stationary position distribution in the full range $-1<a<+1$. Our results for $0<a<+1$ coincide with the known results, while the result for $-1<a<0$ is new. Furthermore, we compute exactly the MFPT in the range $0\leq a < 1$ and show that it increases as $a$ increases from $0$, for any fixed $r$. This indicates that a positive value of the fraction $a$ is not beneficial for the search of a target. For $-1<a<0$, we show that the underlying nonlocal differential equation for the MFPT has a fundamentally different structure compared to the case $0 \leq a <1$. Here we succeeded in finding an exact formula for the MFPT, but only up to an unknown constant that can however be easily inferred from numerical simulations. With this numerically determined constant we then have an exact formula for the MFPT. This solution shows that the MFPT actually decreases when $a$ decreases from $0$, for fixed $r$. This means a negative value of the parameter $-1<a<0$, i.e., when the partial resetting gets coupled with a reflection, is more optimal than the simple resetting ($a=0$). We also performed numerical simulations and found a perfect agreement with our theoretical predictions. 

The rest of this paper is organized as follows. In Section \ref{sec:NESS}, we compute the exact stationary position distribution in the full range $-1<a<1$. In 
Section \ref{sec:MFPT-positive-a}, we compute the exact MFPT in the range $0 \leq a <1$, while in Section \ref{sec:MFPT-negative-a} we study the MFPT for $-1<a<0$. In Section \ref{sec:numerics}, we provide details of numerical simulations. 
Finally, in Section \ref{sec:conclusion} we conclude with a summary and outlook, while some details are relegated to the Appendices.

\begin{figure}
\centering
\includegraphics[width = 0.3\textwidth]{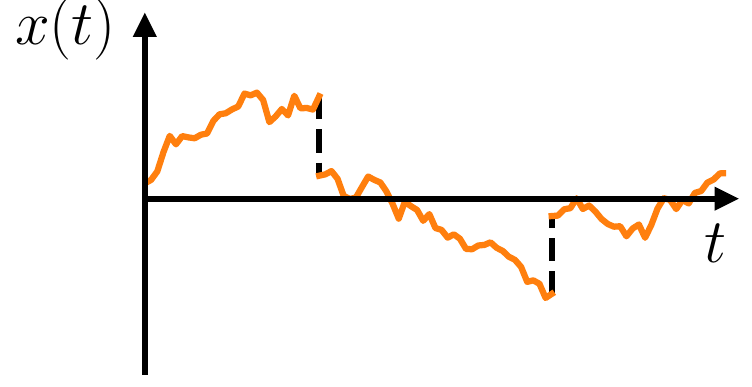}
\hfill
\includegraphics[width = 0.3\textwidth]{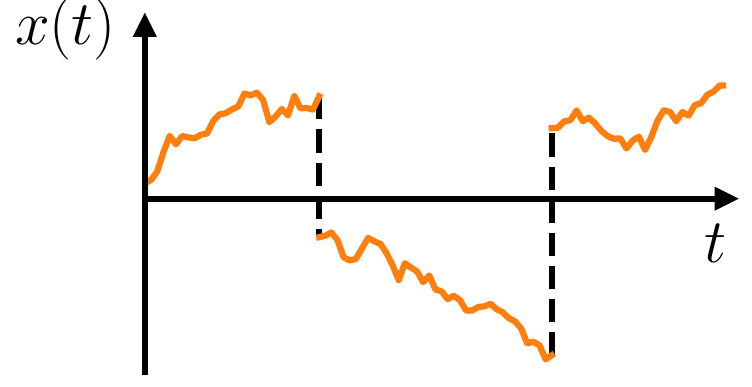}
\hfill
\includegraphics[width = 0.3\textwidth]{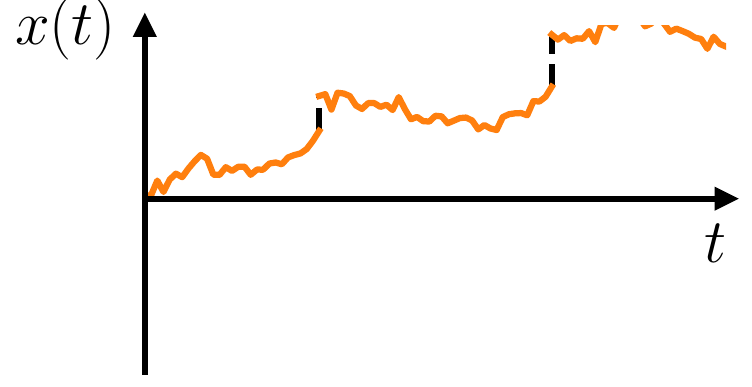}
\caption{Schematic trajectories of a rescaling random walk, as defined in Eq. (\ref{eq:summary-model-dynamics}) for $0 < a < 1$ (left panel), $-1 < a < 0$ (middle panel) and $a > 1$ (right panel). The orange part of the curves represents the purely diffusive part of the trajectory, while the dashed vertical black lines correspond to long-range rescaling (resetting) moves. As shown in the text, the position distribution of the walker becomes stationary only for $-1<a<1$, while it remains non-stationary at all times for $|a| \geq 1$.} \label{fig:model}
\end{figure}

\section{Non-equilibrium steady state} \label{sec:NESS}

We consider the process $x(t)$ evolving via the Langevin equation (\ref{eq:summary-model-dynamics}). Let $P_{r, a}(x, t)$ denote the probability density 
that the particle arrives at $x$ at time $t$, starting from the origin $x = 0$ at time $t = 0$. One can easily deduce the Fokker-Planck equation for $P_{r, a}(x, t)$, valid for arbitrary $a$ as follows. Suppose we increment the time from $t$ to $t +\dd t$. Then the evolution of $P_{r, a}(x, t)$ can be expressed as  
\begin{equation} \label{eq:discrete-balance}
  P_{r, a}(x, t + \dd t) = (1 - r \dd t)
  \Big\langle P_{r, a}(x - \sqrt{2 D \, \dd t}\,\xi(t),t)\Big\rangle_\xi +
  \frac{r \, \dd t}{|a|}  \; P_{r, a}\left( \frac{x}{a}, t \right) \;,
\end{equation}
where the notation $\langle \cdots \rangle_\xi$ means an average over the instantaneous noise $\xi(t)$, which is distributed via a Gaussian with zero mean and unit variance. In Eq. (\ref{eq:discrete-balance}), the first term denotes the diffusive move in time $\dd t$, which occurs with probability $1-r\,\dd t$ (see Eq. (\ref{eq:summary-model-dynamics})). If the particle has to reach $x$ at time $t + \dd t$, it must have been at $x -  \sqrt{2 D \, \dd t}\,\xi(t)$ at time $t$ and one needs to average over all possible values of $\xi(t)$. The second term denotes the resetting move that brings the particle from $x/a$ to $x$ in time $\dd t$. This resetting move with probability $r\, \dd t$ (see Eq. (\ref{eq:summary-model-dynamics})). The factor $1/|a|$ is a Jacobian factor associated to the probability density $P_{r, a}\left( \frac{x}{a}, t \right)$ in $x$. In the limit $\dd t \to 0$, we expand the first term in a Taylor series in $\sqrt{\dd t}$. Keeping terms up to order $O(\dd t)$ and taking the limit $\dd t \to 0$ one arrives at the Fokker-Planck equation
\begin{equation} \label{eq:Fokker-Planck}
\pdv{P_{r, a}(x, t)}{t} = D \pdv[2]{P_{r, a}(x, t)}{x} - r P_{r, a}(x, t) + \frac{r}{|a|} P_{r, a}\left( \frac{x}{a}, t \right) \;.
\end{equation}
This equation is valid for all $a$ and, for $a\geq 0$, it was already derived in Refs.~\cite{TRR22,P22,BCHPM23,H23}. By integrating it over $x$ from $-\infty$ to $+\infty$, one clearly sees that Eq.~(\ref{eq:Fokker-Planck}) conserves the normalization
\bea \label{eq:NESS-normalization}
 \int_{-\infty}^\infty P_{r,a}(x,t)\, \dd x  = 1 \;.
 \eea 
 
 \begin{figure}
\centering
\includegraphics[width=0.6\textwidth]{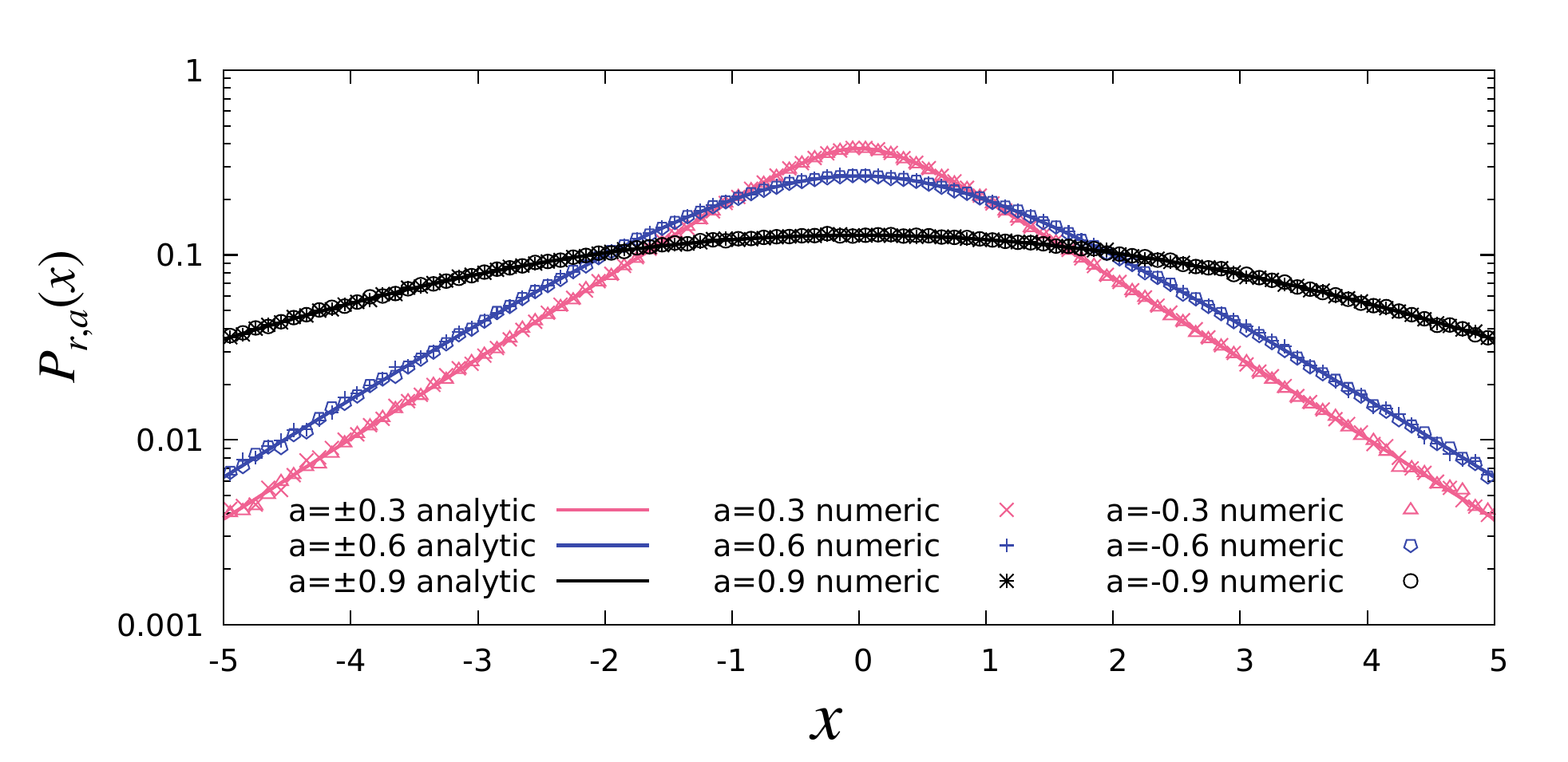}
\caption{Plot of the steady-state probability density $P_{r,a}(x)$ vs $x$ for different values of the parameter $a \in (-1,1)$ and for fixed $r=1$ and $D=1$. 
The analytical result in Eq. (\ref{eq:full-NESS}) is in perfect agreement with the results obtained from numerical simulations.}\label{fig:NESS}
\end{figure}
As mentioned in the introduction, the NESS exists only in the range $|a| < 1$. Assuming that the NESS exists, one can obtain it by setting the left hand side (LHS) of Eq. \eqref{eq:Fokker-Planck} to zero. This gives
\begin{equation} \label{eq:SS-FP}
0 = D \dv[2]{P_{r, a}(x)}{x} - r P_{r, a}(x) + \frac{r}{|a|} P_{r, a}\left( \frac{x}{a}\right) \quad, \quad |a| < 1 \;.
\end{equation}
Even though this is an ordinary differential equation, it is {\it nonlocal} in $x$ and hence is nontrivial to solve. This type of nonlocal equations have appeared in other contexts before, e.g., in the modelling of cell growth \cite{HW89}, in the growth of clusters in a generalized Eden model on a tree \cite{DM06} and also  
in the context of the discrete Ornstein-Uhlenbeck processes~\cite{Lar04,MK07}, etc. These works suggest to look for a solution of (\ref{eq:SS-FP}) in the form 
\begin{equation} \label{eq:NESS-Ansatz}
P_{r, a}(x) = \sum_{n = 0}^{+\infty} c_n e^{-\frac{1}{|a|^n} \sqrt{\frac{r}{D}} |x| } \;,
\end{equation}
where $c_n$ are constants. Substituting this in Eq. (\ref{eq:SS-FP}) we obtain
\begin{equation} \label{eq:SS-Ansatz-ODE}
0 = \sum_{n = 0}^{+\infty} \left( \frac{r}{|a|^{2n}} c_{n} - r c_n + \frac{r}{|a|} c_{n - 1} \right) e^{-\frac{1}{|a|^n} \sqrt{\frac{r}{D}} |x| } \;.
\end{equation}
Since Eq. (\ref{eq:SS-Ansatz-ODE}) must hold for any $x$ the prefactor to the exponential in the series must vanish for any $n$ yielding the recursion relation
\begin{equation} \label{eq:NESS-recursion}
c_n \left(\frac{1}{|a|^{2n}} - 1\right) = -\frac{1}{|a|} c_{n-1} \;.
\end{equation}
Iterating the recursion in Eq. (\ref{eq:NESS-recursion}) we can express any $c_n$ in terms of $c_0$ as
\begin{equation} \label{eq:incomplete-cn}
c_n = \frac{1}{|a|^n} \frac{c_0}{\prod_{k = 1}^n (1 - |a|^{-2k})} \;.
\end{equation}
Using Eq. (\ref{eq:incomplete-cn}) in Eq. (\ref{eq:NESS-Ansatz}) we obtain
\begin{equation} \label{eq:incomplete-NESS}
P_{r, a}(x) = c_0 \left[\sum_{n = 1}^{+\infty} \frac{1}{|a|^n} \frac{1}{\prod_{k = 1}^n (1 - |a|^{-2k})} e^{-\frac{1}{|a|^n} \sqrt{\frac{r}{D}} |x| } + \, e^{- \sqrt{\frac{r}{D}} |x|} \right] \;,
\end{equation}
where we have separated out the $n=0$ term. The only unknown constant $c_0$ is then determined from the normalization condition \eqref{eq:NESS-normalization}, giving
\begin{equation} \label{eq:NESS-c0}
c_0^{-1} = 2 \sqrt{\frac{D}{r}} \left(\sum_{n = 1}^{+\infty} \frac{1}{\prod_{k = 1}^n (1 - |a|^{-2k}) } + 1 \right) \;.
\end{equation}
This gives the exact stationary position distribution 
\begin{equation} \label{eq:full-NESS}
P_{r, a}(x) = \frac{1}{2} \sqrt{\frac{r}{D}} \frac{1}{\left[1 + \sum_{n = 1}^{+\infty} \frac{1}{\prod_{k = 1}^n (1 - |a|^{-2k})}\right]} \left( e^{-\sqrt{\frac{r}{D}} |x| } + \sum_{n = 1}^{+\infty} \frac{1}{|a|^n} \frac{1}{\prod_{k = 1}^n (1 - |a|^{-2k})} e^{-\frac{1}{|a|^n} \sqrt{\frac{r}{D}} |x| } \right) \;.
\end{equation}
The stationary distribution is evidently symmetric around $x=0$ and is plotted in Fig. \ref{fig:NESS} for three different values of $a$, showing a perfect agreement with numerical simulations. In the limit $a \to 0$, we recover the result in Eq. (\ref{P_RBM}), as expected. For $0<a<1$, this solution already appeared in Refs. \cite{P22,BCHPM23}, but for $-1<a<0$, we have not seen this solution in the literature. 

It is interesting to derive the asymptotic behaviors of $P_{r,a}(x)$ for small and large $x$. The large $x$ behavior is simple, since the term containing the sum becomes subleading for large $x$. Hence, we get
\bea \label{Pstat_large}
P_{r, a}(x) \simeq \frac{1}{2} \sqrt{\frac{r}{D}} \frac{1}{\left[1 + \sum_{n = 1}^{+\infty} \frac{1}{\prod_{k = 1}^n (1 - |a|^{-2k})}\right]} \; e^{-\sqrt{\frac{r}{D}} |x| }  \quad, \quad {\rm as} \quad |x| \to \infty \;.
\eea
In contrast, the small $x$ behavior turns out to be much more tricky. From the plot in Fig. \ref{fig:NESS}, it seems to behave as $P_{r,a}(x) \sim P_{r,a}(0) - b x^2$, as $x \to 0$, i.e., the linear term in the small $x$ expansion vanishes. Indeed, by expanding Eq. (\ref{eq:NESS-Ansatz}) up to linear order in $x$, we get
\bea \label{Pstat_small}
P_{r,a}(x) \simeq P_{r,a}(0)- \sqrt{\frac{r}{D}} |x| \sum_{n=0}^\infty \frac{c_n}{|a|^n} \;,
\eea
where the coefficients $c_n$'s are given in Eqs. (\ref{eq:incomplete-cn}) and (\ref{eq:NESS-c0}). For the linear term to vanish, we must have the identity $\sum_{n=0}^\infty c_n/|a|^n = 0$. Using the explicit expressions for $c_n$'s, this amounts to the identity valid for all $0<|a| <1$
\bea \label{identity}
1 + \sum_{n=1}^\infty \frac{1}{|a|^{2n}} \frac{1}{\prod_{k=1}^n (1-|a|^{-2k})} = 0 \;.
\eea
We have numerically checked with Mathematica that it is indeed true for several values of $0<|a|<1$. However, we could not prove this nontrivial identity. For $|a|>1$, the left hand side of Eq. (\ref{identity}) can be expressed as the inverse of the Euler function, however we are not able to find an expression for it for $0<|a|<1$. Proving this identity remains an interesting number theoretical challenge. Thus, to summarize, the stationary distribution behaves as a Gaussian distribution for small $x$, while having an exponential tail for large $x$. Similar asymptotic behaviors for scaling functions also appeared in several models, e.g.,  
in the time-dependent position distribution in models of diffusing diffusivity~\cite{CSMS17,LG18,BB20}, for particles driven by a resetting noise \cite{GMS23} and in certain experimental systems~\cite{WABG09,WGLFGH19}.

\section{Mean first passage time} \label{sec:MFPT}

For a diffusing particle starting at the origin in $d=1$ and resetting stochastically to the origin with rate $r$, we have seen in Eq. (\ref{MFPT_RBM}) that the MFPT $T(0)$ is not only finite, but can also be optimised with respect to $r$. In this section, we study whether the introduction of the additional parameter $a$,  with $|a|<1$, lowers the MFPT compared to the $a=0$ case.

We consider a particle in one-dimension, starting at $x$, and evolving via Eq. (\ref{eq:summary-model-dynamics}), with a target located at $x=L$. Our goal is to compute the MFPT $T_{r,a,L}(x)$ to find the target at $L$, starting at $x$.  
Eventually, for simplicity, we will focus on the case $x=0$, but for the moment we keep $x$ arbitrary, since we will derive a backward Fokker-Planck type differential equation for $T_{r,a,L}(x)$, with $x$ as a variable. To derive this equation, it is convenient to start with $S(x,t)$ denoting the survival probability  
of the target up to time $t$, i.e., the probability that the target is not found by the particle up to time $t$. Consequently $F(x,t) = - \partial_t S(x,t)$ denotes the first-passage time distribution to the target. The MFPT is just the first moment of $F(x,t)$, i.e., 
\begin{equation} \label{eq:MFPT-integral}
T_{r,a,L}(x) = \int_0^{+\infty} \dd t \; \left(- \pdv{S(x, t)}{t} \right) t = \int_0^{+\infty} S(x, t) \dd t \;,
\end{equation}
where we performed an integration by parts and assumed that $S(x, t) t \to 0$ as $t \to +\infty$ which can be checked a posteriori.
For the brevity of notations, we will omit the subscripts of the MFPT $T(x) \equiv T_{r, a, L}(x)$. We now consider the backward evolution equation
for $S(x,t)$. We consider a trajectory of duration $t+\dd t$, starting at $x$ and evolving via Eq. (\ref{eq:summary-model-dynamics}). We split the interval $[0, t + \dd t]$ into two intervals $[0, \dd t]$ and $[\dd t, t+\dd t]$. During the first interval, the walker diffuses to a new position $x + \sqrt{2 D \, \dd t} \, \xi(0)$ with probability $1-r\,\dd t$ and with the complementary probability $r\,\dd t$, it resets to a new position $a\,x$. Here $\xi(0)$ denotes the initial random jump. 
During the second interval, the evolution proceeds starting at the new position at the end of the first interval $[0, \dd t]$. Consequently, we can write 
\begin{equation} \label{eq:BFP}
S(x, t + \dd t) = (1 - r \dd t)  \Big\langle S(x + \sqrt{2 D \, \dd t}\,\xi(0))\Big\rangle_\xi + {r \, \dd t}  \; S\left( {a} \,x, t \right) \;.
\end{equation}
Taking the limit $\dd t\to 0$, one arrives at the backward equation 
\begin{equation} \label{eq:survival-FP}
\pdv{S(x, t)}{t} = D \pdv[2]{S(x, t)}{x} - r S(x, t) + r S(a x, t) \;.
\end{equation}
This equation is valid in the range $x \in (-\infty, +\infty)$ with the absorbing boundary condition 
\bea \label{bc_S}
S(x=L,t) = 0 \quad, \quad {\rm for \; all} \quad  t \geq 0 \;.
\eea
This condition comes from the fact that if the particle starts exactly at $x=L$, it immediately finds the target and hence the survival
probability vanishes. Furthermore, as $x \to \pm \infty$, the survival probability must remain upper bounded by unity. The initial condition reads
\bea \label{ic_S}
S(x,t=0) = 1 \quad {\rm for \; all \;} \quad x \neq  L \;.
\eea
Integrating Eq. (\ref{eq:survival-FP}) over $t$ from $0$ to $\infty$ and using the initial condition (\ref{ic_S}), we get, using Eq. (\ref{eq:MFPT-integral}), the backward differential equation for $T(x)$
\begin{equation} \label{eq:MFPT-ODE}
D T''(x) - r T(x) + r T(a x) = - 1\;,
\end{equation}
valid for $x \in (-\infty, +\infty)$ with the absorbing boundary condition 
\bea \label{bc_T}
T(L) = 0 \;.
\eea
In addition, as $x \to \pm \infty$, the MFPT $T(x)$ can not diverge faster than $\sim x^2$, since
diffusion is the slowest mode of transport and the resetting can only reduce the MFPT. Once again, this ordinary second-order equation (\ref{eq:MFPT-ODE}) is nonlocal in $x$, making it nontrivial to solve. We will see below that the solution is actually very different for $0\leq a<1$ and $-1<a \leq 0$. We discuss the two cases separately in the two subsections below.

\subsection{Positive rescaling: $0 \leq a < 1$} \label{sec:MFPT-positive-a}


In this subsection, our goal is to calculate the MFPT $T(0)$ starting from the origin for the case $0 \leq a < 1$. To compute this, we need to solve the nonlocal 
backward differential equation (\ref{eq:MFPT-ODE}) with $T(x)$ denoting the MFPT starting from the initial position $x \leq L$. Upon finding the solution for $T(x)$ for arbitrary $x \leq L$, we will eventually set $x=0$. Since $x=0 \leq L$, we need to solve the differential equation only in the region $x \leq L$ and henceforth we will not consider the case $x > L$. Note that the nonlocal term in Eq. (\ref{eq:MFPT-ODE}) involves the location $a\,x$ which always stays to the left of $L$. Hence, the particle never jumps to the right of $L$ and we just need to solve the differential equation for $x \leq L$. Since there is no known general method to solve such nonlocal equations, we try below a power series solution in $x$ and show that it leads to an exact solution. We substitute
\begin{equation} \label{eq:MFPT-power-series}
T(x) = \sum_{n = 0}^{+\infty} b_n x^n 
\end{equation}
in Eq. (\ref{eq:MFPT-ODE}) and solve recursively for the $b_n$'s. This gives 
\begin{equation} \label{eq:MFPT-power-series-ODE}
-1 = \sum_{n = 0}^{+\infty} \left[D b_{n+2} (n+2)(n+1) - r b_n + r b_n a^n \right] x^n \;.
\end{equation}
This equation holds for any $x \leq L$. Therefore setting $x = 0$ sets the first even constant
\begin{equation} \label{eq:MFPT-c2}
-1 = 2 D b_2 \;.
\end{equation}
Then the remaining terms of the series must all vanish which leads to the following recursion relation
\begin{equation} \label{eq:MFPT-c-recursion}
b_{n+2} D (n+2) (n+1) = r (1 - a^n)b_n \;.
\end{equation}
Iterating Eq. (\ref{eq:MFPT-c-recursion}) we can express any $b_n$ for $n \geq 1$ as a function of $b_1$ and $b_2$
\begin{equation} \label{eq:MFPT-c}
b_{2n} = \frac{2 b_2}{(2n)!} \left(\frac{r}{D}\right)^{n-1} \prod_{j = 1}^{n-1} (1 - a^{2j}) \mbox{~~and~~} b_{2n+1} = \frac{b_1}{(2n+1)!} \left( \frac{r}{D} \right)^n \prod_{j = 0}^{n-1} (1 - a^{2j+1})  \;.
\end{equation}
Hence the only constants which are still undetermined are $b_1$ and $b_0$. Putting Eqs. (\ref{eq:MFPT-c2}), (\ref{eq:MFPT-c}) and (\ref{eq:MFPT-power-series}) together we obtain
\begin{equation} \label{eq:MFPT-incomplete}
T(x) = b_0 + b_1 \sqrt{\frac{D}{r}} \sum_{n = 0}^{+\infty} \frac{1}{(2n + 1)!} \left(\sqrt{\frac{r}{D}} x\right)^{2n + 1} \prod_{j = 0}^{n-1} (1 - a^{2j+1}) - \frac{1}{r} \sum_{n = 1}^{+\infty} \frac{1}{(2n)!} \left( \sqrt{\frac{r}{D}} x \right)^{2n} \prod_{j = 1}^{n-1} (1 - a^{2j}) \;.
\end{equation}
Note that in the $n=0$ term of the sum multiplying $b_1$, the product is interpreted as unity. Using the absorbing boundary condition in Eq. (\ref{bc_T}) gives the first relation 
\begin{equation} \label{eq:MFPT-c0}
b_0 = - b_1 \sqrt{\frac{D}{r}} \sum_{n = 0}^{+\infty} \frac{1}{(2n + 1)!} \left(\sqrt{\frac{r}{D}} L\right)^{2n + 1} \prod_{j = 0}^{n-1} (1 - a^{2j+1}) + \frac{1}{r} \sum_{n = 1}^{+\infty} \frac{1}{(2n)!} \left( \sqrt{\frac{r}{D}} L \right)^{2n} \prod_{j = 1}^{n-1} (1 - a^{2j}) \;.
\end{equation}
Using Eq. (\ref{eq:MFPT-c0}) we can re-write Eq. (\ref{eq:MFPT-incomplete}) as
\begin{align}
T(x) = b_1 \sqrt{\frac{D}{r}} \sum_{n = 0}^{+\infty} \frac{1}{(2n + 1)!}  \left(\sqrt{\frac{r}{D}} \right)^{2n+1} (x^{2n+1} - L^{2n+1}) \prod_{j = 0}^{n-1} (1 - a^{2j+1}) \nonumber\\
- \frac{1}{r} \sum_{n = 1}^{+\infty} \frac{1}{(2n)!} \left(\sqrt{\frac{r}{D}} \right)^{2n} (x^{2n} - L^{2n})  \prod_{j = 1}^{n-1} (1 - a^{2j}) \;. \label{eq:MFPT-incomplete-2}
\end{align}
To simplify these expressions, we introduce the following compact notations
\begin{equation} \label{eq:def-Cn}
C_1 = 1, \quad C_2 = 1, \quad C_{2n} = \prod_{j = 1}^{n-1} (1 - a^{2j}) \mbox{~~for~~} n \geq 2, \quad  C_{2n + 1} = \prod_{j = 0}^{n-1} (1 - a^{2j+1}) \mbox{~~for~~} n \geq 1 \; 
\end{equation}
and rescaled distances
\bea \label{def_beta}
\beta = L \sqrt{\frac{r}{D}} \quad, \quad y = x  \sqrt{\frac{r}{D}} \;.
\eea
In terms of these rescaled quantities, Eq. (\ref{eq:MFPT-incomplete-2}) simplifies to
\begin{align}
T(x) = b_1 \sqrt{\frac{D}{r}} \sum_{n = 0}^{+\infty} \frac{C_{2n+1}}{(2n + 1)!}  \left(y^{2n+1} - \beta^{2n+1}\right)  - \frac{1}{r} \sum_{n = 1}^{+\infty} \frac{C_{2n}}{(2n)!} \left(y^{2n} - \beta^{2n}\right)  \;. \label{T_compact}
\end{align}
We have already used the absorbing boundary condition at $x=L$. To fix the only unknown constant $b_1$, we need to investigate the other boundary when $x \to - \infty$. As mentioned earlier, the MFPT $T(x)$ should not grow faster than $\sim x^2$ as $x \to - \infty$. We now show that this condition uniquely fixes the unknown constant $b_1$. 
\begin{figure}[t]
\includegraphics[width = 0.5\textwidth]{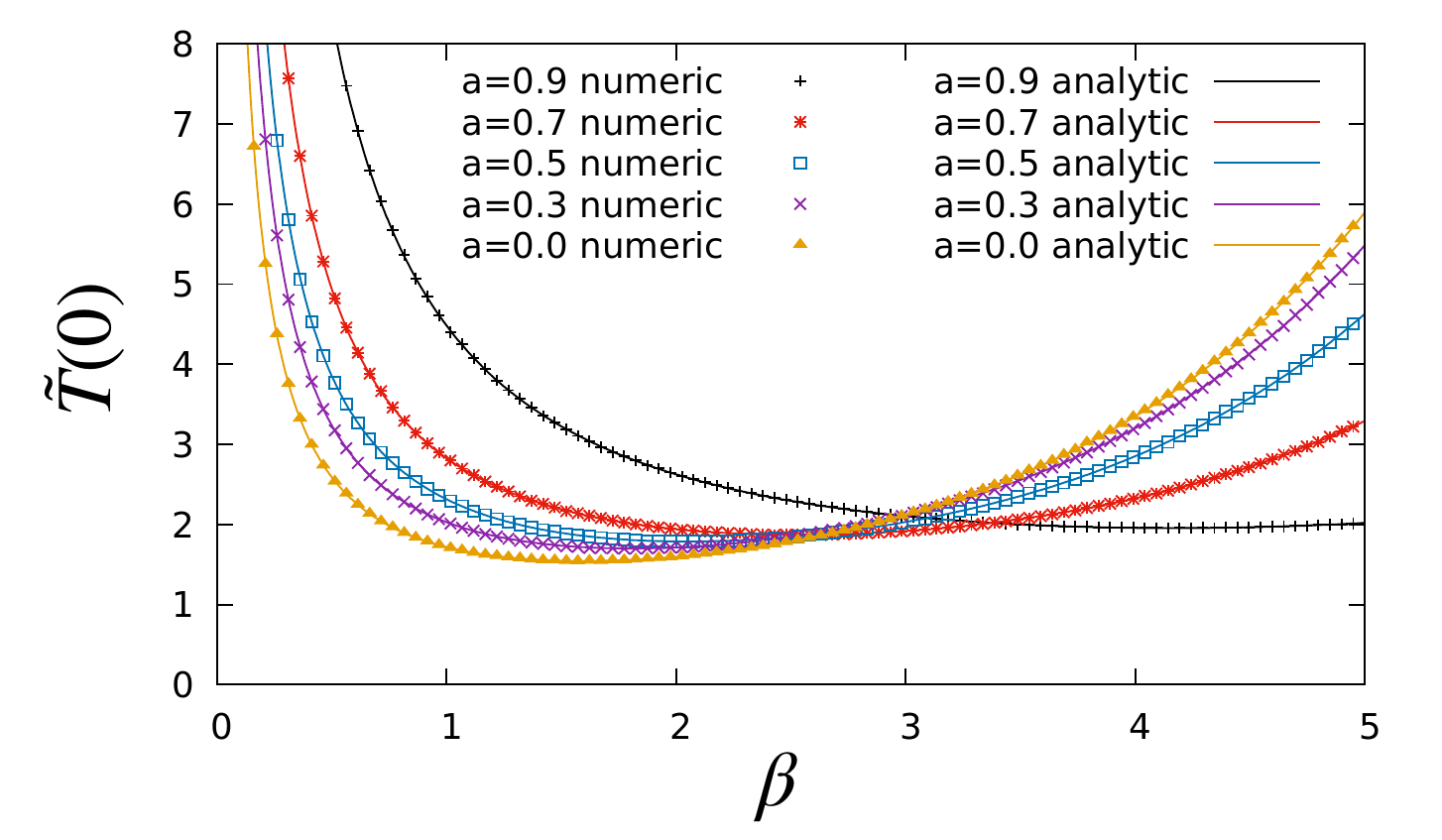}
\caption{The dimensionless MPFT $\tilde T(0)$ as a function of $\beta = \sqrt{r/D} L$ for different values of $0 \leq a < 1$. The analytical result in Eq. (\ref{eq:MFPT-pos-a-dimensionless}) is in excellent agreement with the results of numerical simulations for different values of $0 \leq a < 1$. For a fixed $a$, the MFPT $\tilde T(0)$, as a function of $\beta$, has a minimum at $\beta = \beta^*(a)$.}\label{Ttilde_apos}
\end{figure}

To analyse $T(x)$ in Eq. (\ref{T_compact}) in the limit $x \to -\infty$, we first note that $\left(y^{2n+1} - \beta^{2n+1}\right)  \sim y^{2n+1}$ and similarly $\left(y^{2n} - \beta^{2n}\right) \sim y^{2n}$. Furthermore, for large $|y|$ both sums in Eq. (\ref{T_compact}) are dominated by large $n$. Hence, we can replace $C_{2n+1}$ and $C_{2n}$ by their respective $n \to \infty$ limits, i.e., 
\bea \label{lim_C}
\lim_{n \to \infty} C_{2n+1} = \prod_{j=0}^\infty (1-a^{2j+1}) \quad, \quad \lim_{n \to \infty} C_{2n} = \prod_{j=1}^\infty (1-a^{2j}) \;.
\eea
Therefore, to leading order for large $|y|$, we can take out the $C$-factors outside the sums and evaluate the sums explicitly to get
\bea \label{T_large_y}
T(x) \underset{x \to -\infty}{\simeq} - b_1 \sqrt{\frac{D}{r}}  \left[\prod_{j=0}^\infty (1-a^{2j+1})\right] \sinh{(y)} + \frac{1}{r} \left[\prod_{j=1}^\infty (1-a^{2j})\right]\left( \cosh{(y)}-1\right) \;.
\eea 
Thus, to leading order for large $|x|$, this expression diverges as $T(x) \sim \exp{|y|} =  \exp(\sqrt{r/D}|x|)$. Since this is not allowed physically, the amplitude of this term must vanish. This fixes the constant $b_1$ uniquely as
\begin{equation}\label{eq:MFPT-c1}
-b_1 \sqrt{\frac{D}{r}} = \frac{1}{r} \frac{\prod_{j = 1}^{+\infty} (1 - a^{2j})}{\prod_{j = 0}^{+\infty} (1 - a^{2j +1})} = \frac{1}{r}\, R_a \;.
\end{equation}
where $R_a$ is given by the ratio
\begin{equation} \label{eq:def-Ra}
R(a) = \frac{\lim_{n\to+\infty} C_{2n}}{\lim_{n \to +\infty} C_{2n +1}} = \frac{\prod_{j = 1}^{+\infty} (1 - a^{2j})}{\prod_{j = 0}^{+\infty} (1 - a^{2j +1})} \;.
\end{equation}
Thus the final expression for the MFPT $T(x)$, in terms of the rescaled coordinates (\ref{def_beta}), reads 
\begin{equation} \label{eq:MFPT-pos-a}
T\left(x = y \sqrt{\frac{D}{r}}\right) = \frac{1}{r} \Bigg[R(a) \sum_{n = 0}^{+\infty} \frac{C_{2n + 1}}{(2n +1)!} (\beta^{2n+1} - y^{2n+1}) + \sum_{n = 1}^{+\infty} \frac{C_{2n}}{(2n)!}  (\beta^{2n} - y^{2n})  \Bigg] \;.
\end{equation}
Setting $x = 0$ we obtain the MFPT for a random walk starting at the origin to reach a target at $L$ 
\begin{equation} \label{eq:MFPT-pos-a-origin}
T(0) = \frac{1}{r} \left[ \sum_{n = 1}^{+\infty} \frac{C_{2n}}{(2n)!}\beta^{2n} + R(a) \sum_{n = 0}^{+\infty} \frac{C_{2n+1}}{(2n+1)!} \beta^{2n+1} \right] \;,
\end{equation}
with $R(a)$ defined in Eq. (\ref{eq:def-Ra}). Furthermore, one can also define a dimensionless MFPT $\tilde{T}(0) = D\,T(0)/L^2$, that depends only on two dimensionless parameters: $a$ and $\beta = L \sqrt{r/D}$, which reads
\begin{equation}\label{eq:MFPT-pos-a-dimensionless}
\tilde{T}(0) = \frac{1}{\beta^2} \left[\sum_{n = 1}^{+\infty} \frac{C_{2n} \beta^{2n} }{(2n)!} + R(a) \sum_{n = 0}^{+\infty} \frac{C_{2n+1} \beta^{2n+1}}{(2n+1)!}\right] \;. 
\end{equation}
\begin{figure}[t]
\includegraphics[width = 0.9\textwidth]{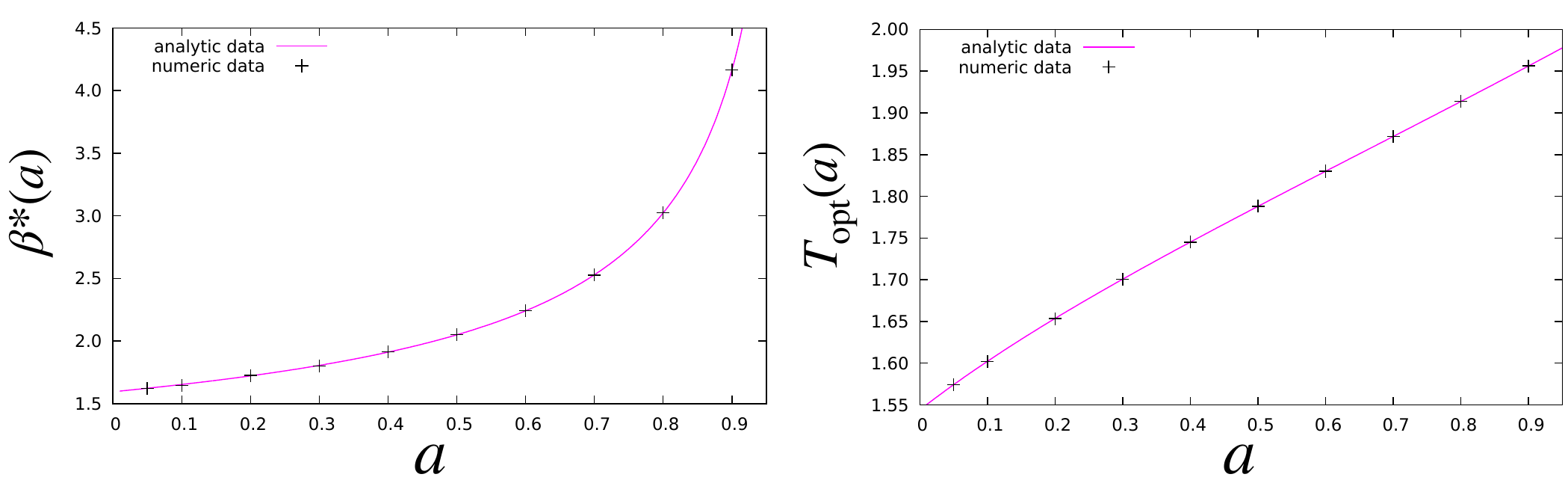}
\caption{{\bf Left:} The optimal value $\beta^*(a)$, at which $\tilde T(0)$ achieves its minimum as a function of $\beta$ for a fixed $a$, plotted as a function of $a$ for $0 \leq a < 1$. {\bf Right:} The optimal MFPT $T_{\rm opt}(a)$, i.e., $\tilde T(0)$ evaluated at $\beta = \beta^*(a)$, plotted as a function of $0 \leq a < 1$.}\label{Figbetastarapos}
\end{figure}
For fixed $a$, the asymptotic behaviors of $\tilde{T}(0)$ for small and large $\beta$ are given by 
\begin{equation} \label{eq:MFPT-pos-a-beta-asymptotics}
\tilde{T}(0) \simeq \begin{dcases}
R(a)/\beta &\mbox{~~when~~} \beta \ll 1 \\
\frac{e^\beta}{\beta^2} \prod_{j = 1}^{+\infty} (1 - a^{2j}) &\mbox{~~when~~} \beta \gg 1 \;.
\end{dcases}
\end{equation}
The small $\beta$ asymptotics is easy to obtain from Eq. (\ref{eq:MFPT-pos-a-dimensionless}) where the $n=0$ term contributes to the dominant order for small $\beta$, leading to the first line in Eq. (\ref{eq:MFPT-pos-a-beta-asymptotics}). To derive the large $\beta$ behavior in the second line of Eq. (\ref{eq:MFPT-pos-a-beta-asymptotics}), we note that the sums in Eq. (\ref{eq:MFPT-pos-a-beta-asymptotics}) are dominated by the large values of $n$ where both $C_{2n}$ and $C_{2n+1}$ converge to the asymptotic values [see Eq. (\ref{eq:def-Cn})]
\bea \label{conv_C}
C_{2n} \underset{n \to \infty}{\longrightarrow} \prod_{j=1}^\infty (1-a^{2j}) \quad, \quad C_{2n+1} \underset{n \to \infty}{\longrightarrow} \prod_{j=0}^\infty (1-a^{2j+1}) \;.
\eea 
Substituting these behaviors in Eq. (\ref{eq:MFPT-pos-a-dimensionless}) and using \eqref{eq:MFPT-pos-a-beta-asymptotics}, we get the second line of Eq. (\ref{eq:MFPT-pos-a-beta-asymptotics}). Thus we see that, as a function of $\beta$ for fixed $a$, the MFPT $\tilde T(0)$ diverges in both limits $\beta \to 0$ and $\beta \to \infty$. Thus, it indicates that it may have a unique minimum at $\beta = \beta^*(a)$. Indeed, the analytical result in Eq. (\ref{eq:MFPT-pos-a-dimensionless}) can easily be plotted and it shows a unique minimum (see Fig. \ref{Ttilde_apos}). This minimum $\beta^*(a)$ can easily be determined by setting the derivative of Eq. (\ref{eq:MFPT-pos-a-dimensionless}) with respect to $\beta$ to zero and determining the root using Mathematica. This optimal value $\beta^*(a)$, as a function of $a$, is shown in the left panel of Fig. \ref{Figbetastarapos}. Finally, the optimal value $T_{\rm opt}(a)$ of the MFPT, i.e., $\tilde T(0)$ evaluated at $\beta = \beta^*(a)$, is plotted as a function of $a$ in the right panel of Fig. \ref{Figbetastarapos}. Clearly, one sees that $T_{\rm opt}(a)$ is a monotonically increasing function of $a$ in the range $a \in [0,1]$.

\vspace*{0.5cm}

\noindent {\bf Small $a$ expansion.}  For small $a$, one can make an explicit analysis. Expanding Eq. (\ref{eq:MFPT-pos-a-dimensionless}) to linear order for small $a$, we get
\begin{equation} \label{eq:MFPT-pos-a-a-asymptotics}
\tilde{T}(0) \simeq \frac{1}{\beta^2} \left[e^\beta - 1  + a \beta + O(a^2)\right] \mbox{~~when~~} a \ll 1 \;.
\end{equation}
Note that for $a=0$, we perfectly recover the known result stated in Eq. (\ref{MFPT_RBM}). In fact, this result in Eq. (\ref{eq:MFPT-pos-a-a-asymptotics}) for small $a$ can also be derived directly from the differential equation (\ref{eq:MFPT-ODE}) by making an expansion for small $a$, as shown in Appendix \ref{App_smalla}. Thus, for fixed $\beta$, as $a$ increases from $0$, the MFPT also increases. Taking a derivative of Eq. (\ref{eq:MFPT-pos-a-a-asymptotics}) with respect to $\beta$ and setting it to zero, we get the optimal $\beta^*(a)$ to linear order in $a$
\bea \label{betastar_small_a}
\beta^*(a) = \beta^*(0) + a \frac{\beta^*(0)}{\beta^*(0)-1}\,e^{-\beta^*(0)} + O(a^2) = 1.59362\ldots + a\, 0.54547\ldots  + O(a^2)\;,
\eea
where $\beta^*(0) = 1.59362\ldots$ is the optimal value of $\beta$ for $a=0$. Thus as $a$ increases from $0$, the optimal $\beta^*(a)$ increases for small $a$, which is consistent with the result shown in the left panel of Fig. \ref{Figbetastarapos}. Subsequently, the optimal MFPT $T_{\rm opt}(a)$, up to linear order in $a$, reads
\bea \label{Topt_small_a}
T_{\rm opt}(a) = T_{\rm opt}(0) + a\, \frac{[\beta^*(0)]^2 - 2 (1-e^{-\beta^*(0)})}{[\beta^*(0)]^2(\beta^*(0)-1)} + O(a^2) = 1.54413\ldots + a\, 0.627500\ldots  + O(a^2)\;.
\eea
Therefore, as $a$ increases from $0$, the optimal MFPT increases linearly with $a$ for small $a$, consistent with the results reported in the right panel of Fig. \ref{Figbetastarapos}. 

Thus, in conclusion, introducing a positive value of the rescaling parameter $a$ does not improve the efficiency of the search process, since $T_{\rm opt}(a)$ increases with an increasing positive $a$. In the next subsection, we will see that the situation is drastically different for $-1<a\leq 0$. In this case, 
the optimal MFPT with $-1<a\leq 0$ is {\it lower} than its value at $a=0$.


\subsection{Negative rescaling: $-1 < a \leq 0$} \label{sec:MFPT-negative-a}


\begin{figure}[t]
\includegraphics[width = 0.7 \linewidth]{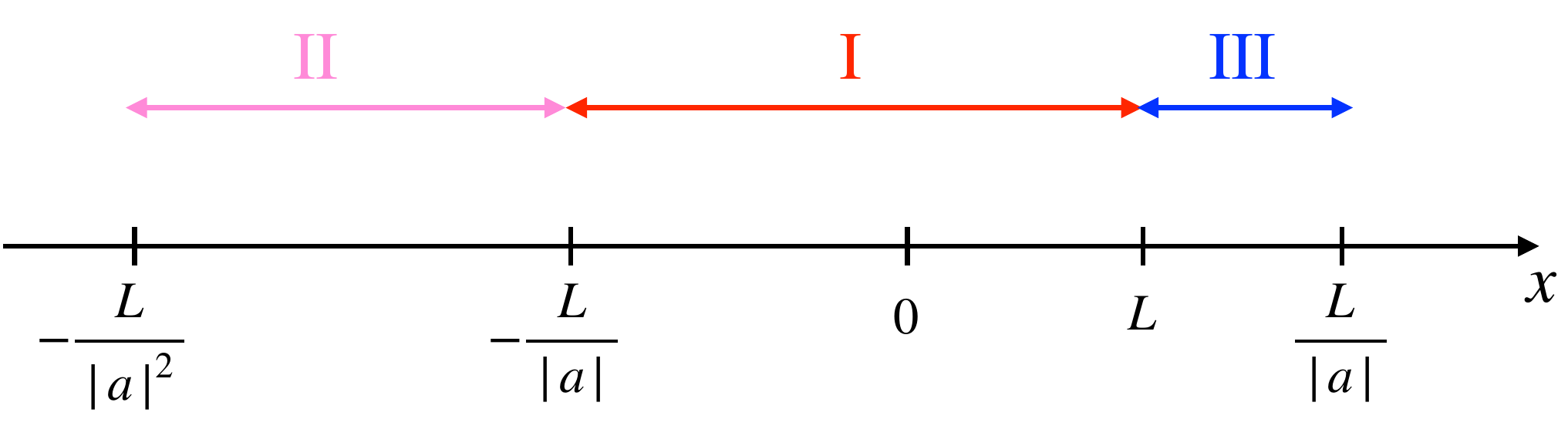}
\caption{For $-1< a \leq 0$, the nonlocal differential equation (\ref{eq:MFPT-ODE_aneg}) needs to be solved in different segments that are interconnected. The segment ${\rm I}$ denotes the region $x \in [-L/|a|,L]$, the segment ${\rm II}$ denotes the region $[-L/|a|^2, -L/|a|]$ and the segment ${\rm III}$ denotes the region $[L,L/|a|]$.}\label{Figa<0}
\end{figure}

In this subsection we consider the complementary case $-1<a\leq 0$. Once again, we need to solve the differential equation (\ref{eq:MFPT-ODE}) for $T(x)$ and eventually set $x=0$. However, contrary to the case $0 \leq a <1$ discussed in the previous subsection, for $-1< a \leq 0$, solving the nonlocal differential equation (\ref{eq:MFPT-ODE}) is much more complicated. To see why, we first rewrite Eq. (\ref{eq:MFPT-ODE}) for $-1<a\leq 0$ as
\begin{equation} \label{eq:MFPT-ODE_aneg}
-1 = D T''(x) - r T(x) + r T(-|a| x) \;.
\end{equation}
We still have the boundary conditions $T(L) = 0$ and the fact that $T(x)$ should not grow faster $\sim |x|^2$ as $x \to \pm \infty$. To solve this nonlocal equation \eqref{eq:MFPT-ODE_aneg} at a given point $x$, the source term $ r T(-|a| x)$ comes from the point $-|a| x$. Consequently, one needs to separate the full line into different segments, as shown in Fig. \ref{Figa<0}. First, we note that, if $-L/|a|\leq x \leq L$ (segment I in Fig. \ref{Figa<0}), the source point $-|a| x$ also belongs to this segment. Thus, the segment I closes onto itself. However, we have only one known boundary condition $T(L) = 0$ and have no information on the MFPT at the other edge at $x = - L/|a|$. For the moment, let us denote this value by $\kappa >0$. In other words, the Eq. (\ref{eq:MFPT-ODE_aneg}) in segment I satisfies the boundary conditions
\begin{equation} \label{eq:MFPT-boundary-neg}
T(L) = 0 \mbox{~~and~~} T(-L/|a|) = \kappa \;,
\end{equation}
where $\kappa$ is unknown. To determine $\kappa$, we need to solve Eq. \eqref{eq:MFPT-ODE_aneg}  in segment II, where $-L/|a|^2 \leq x \leq -L/|a|$. However, for $x$ belonging to this segment II, the source point $-|a| x$ belongs to the segment III in Fig. \ref{Figa<0} where $x \in [L, L/|a|]$. Thus the solution in segment II requires the solution from segment III and this mechanism continues till one arrives at $x = \pm \infty$. Thus this breaks the whole line into different segments. In Fig. \ref{Figa<0}, for simplicity, we show only three of them. Hence we have to iteratively solve (\ref{eq:MFPT-ODE_aneg}) in each segment in order to use the boundary condition as $x \to \pm \infty$. This makes the problem much more complicated to solve. However we see that the segment I is ``closed'' onto itself (i.e., it does not involve other segments), and the full function $T(x)$ for $-L/|a| \leq x \leq L$ can be fully determined, but up to only one unknown constant $\kappa$ representing the MFPT at $-L/|a|$. Hence our strategy would be to first analytically solve Eq. (\ref{eq:MFPT-ODE_aneg}) in segment I, i.e., for $-L/|a| \leq x \leq L$ with $\kappa$ as a given parameter and then use the value of $\kappa$ obtained from numerical simulations. This will fully determine $T(x)$ in segment I and since the origin belongs to that segment, we can set $x=0$ to find $T(0)$. Below, we derive the solution $T(x)$ in terms of $\kappa$.    

We note that to solve $T(x)$ in segment I, i.e., for $-L/|a| \leq x \leq 0$, with $-1<a\leq 0$, the procedure is identical as in the previous subsection, i.e., we try a series expansion as in Eq. (\ref{eq:MFPT-power-series}). Following exactly the same steps that led to the result in Eq. (\ref{eq:MFPT-incomplete-2}), we obtain
\begin{align}
T(x) = b_1 \sqrt{\frac{D}{r}} \sum_{n = 0}^{+\infty} \frac{1}{(2n + 1)!}  \left(\sqrt{\frac{r}{D}} \right)^{2n+1} (x^{2n+1} - L^{2n+1}) \prod_{j = 0}^{n-1} (1 + |a|^{2j+1}) \nonumber\\
- \frac{1}{r} \sum_{n = 1}^{+\infty} \frac{1}{(2n)!} \left(\sqrt{\frac{r}{D}} \right)^{2n} (x^{2n} - L^{2n})  \prod_{j = 1}^{n-1} (1 - |a|^{2j})  \label{eq:MFPT-aneg} \;,
\end{align}
where we used the boundary condition $T(x=L)=0$. We now use the other boundary condition $T(-L/|a|) = \kappa$. This gives 
\begin{equation} \label{eq:kappa}
\kappa = T(-L/|a|) = b_1 \sqrt{\frac{D}{r}} \sum_{n = 0}^{+\infty} \frac{C_{2n + 1}}{(2n+1)!} \left(\sqrt{\frac{r}{D}} L \right)^{2n+1} (-1/|a|^{2n+1} - 1) - \frac{1}{r} \sum_{n = 1}^{+\infty} \frac{C_{2n}}{(2n)!} \left(\sqrt{\frac{r}{D}} L \right)^{2n} (1/|a|^{2n} - 1) \;,
\end{equation}
where the coefficients $C_n$'s are the same as in Eq. \eqref{eq:def-Cn}. The unknown constant $b_1$ is then determined by solving Eq.~\eqref{eq:kappa}. To express this solution in a more compact form, it is convenient to introduce two functions
\begin{equation} \label{eq:def-feven-fodd}
f_{\rm even}(x) = \sum_{n = 1}^{+\infty} \frac{C_{2n}}{(2n)!} x^{2n} \quad \mbox{~~and~~} \quad f_{\rm odd}(x) = \sum_{n = 0}^{+\infty} \frac{C_{2n+1}}{(2n+1)!} x^{2n+1} \;.
\end{equation}
Then solving Eq. (\ref{eq:kappa}) for $b_1$ we get
\begin{equation} \label{eq:c1-neg-a}
b_1 = -\sqrt{\frac{r}{D}}\, \left[  \frac{r \kappa + f_{\rm even}\left(\sqrt{\frac{r}{D}} \frac{L}{|a|}\right) - f_{\rm even}\left(\sqrt{\frac{r}{D}} L\right)}{  r f_{\rm odd}\left( \sqrt{\frac{r}{D}} \frac{L}{|a|} \right) + r f_{\rm odd}\left( \sqrt{\frac{r}{D}} L \right) } \right]  \;.
\end{equation}

\begin{figure}[t]
\includegraphics[width = 0.5\textwidth]{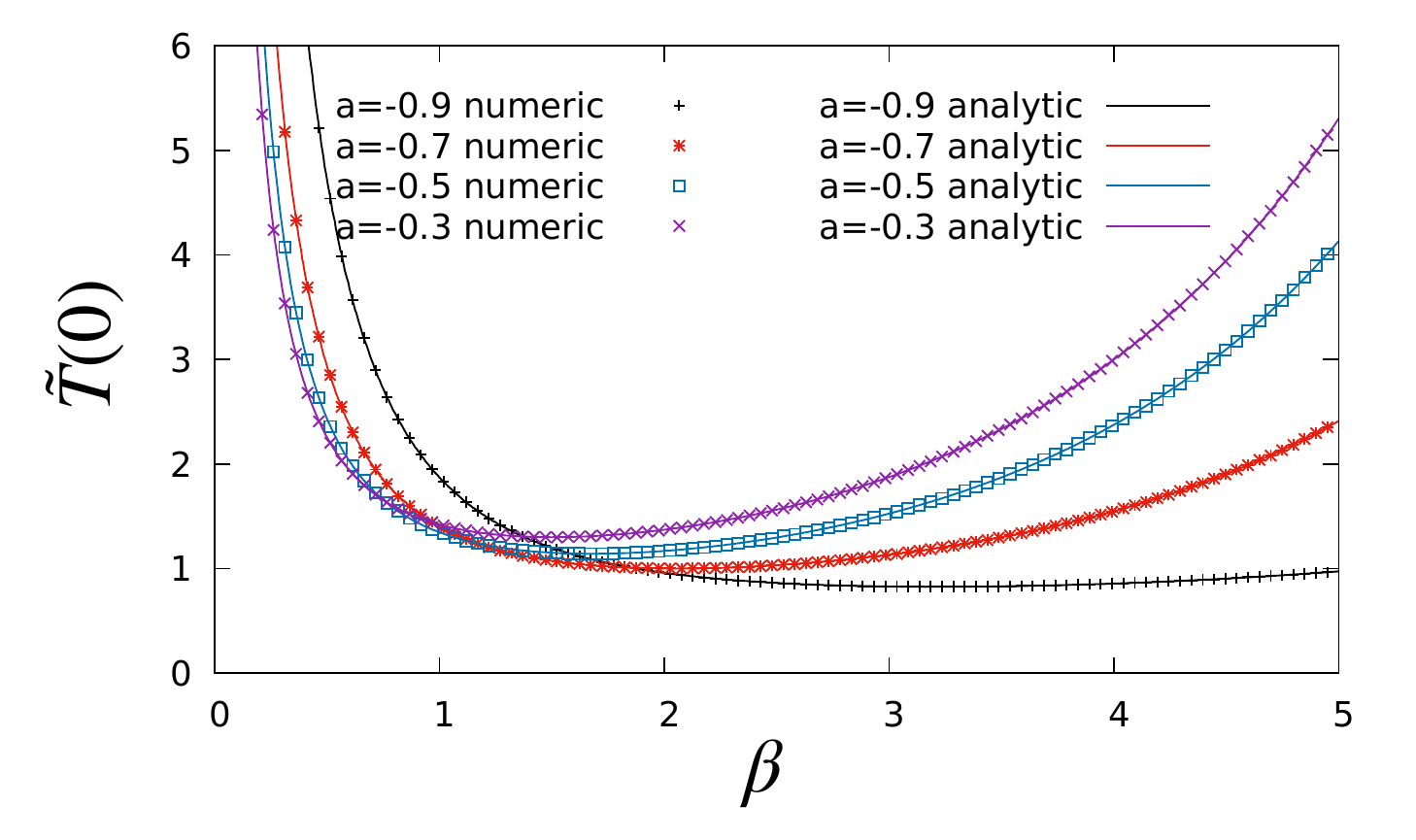}
\caption{The dimensionless MPFT $\tilde T(0)$ as a function of $\beta = \sqrt{r/D} L$ for different values of $-1< a \leq 0$. The analytical result in Eq. (\ref{eq:MFPT-neg-a-origin-dimensionless}) (with $\tilde T(-L/|a|)$ taken as an input from numerical simulations) is in excellent agreement with numerical simulations for $\tilde T(0)$. For a fixed $a$, the MFPT $\tilde T(0)$, as a function of $\beta$, has a minimum at $\beta = \beta^*(a)$.}\label{Ttilde_aneg}
\end{figure}
Substituting this expression for $b_1$ in Eq. (\ref{eq:MFPT-aneg}) we obtain the full MFPT $T(x)$ in segment I (i.e., for $x \in [-L/|a|,L]$) for the case $-1 < a \leq 0$ as
\begin{align}
T(x) = \frac{f_{\rm even}\left(\sqrt{\frac{r}{D}} L\right) - f_{\rm even}\left(\sqrt{\frac{r}{D}} \frac{L}{|a|}\right) - r \kappa }{ r f_{\rm odd}\left( \sqrt{\frac{r}{D}} \frac{L}{|a|} \right) + r f_{\rm odd}\left( \sqrt{\frac{r}{D}} L \right) } \left[ f_{\rm odd}\left(\sqrt{\frac{r}{D}} x\right) - f_{\rm odd}\left(\sqrt{\frac{r}{D}} L\right) \right] \nonumber\\
- \frac{1}{r} \left[ f_{\rm even}\left( \sqrt{\frac{r}{D}} x \right) - f_{\rm even}\left( \sqrt{\frac{r}{D}} L \right) \right] \;. \label{eq:MFPT-neg-a}
\end{align}
Setting $x = 0$ we obtain the MFPT for a walker starting from the origin to reach a target at $L$ 
\begin{equation} \label{eq:MFPT-neg-a-origin}
T(0) =  \frac{1}{r} \frac{1}{f_{\rm odd}\left( \sqrt{\frac{r}{D}} \frac{L}{|a|} \right) + f_{\rm odd}\left( \sqrt{\frac{r}{D}} L \right) } \left[ f_{\rm even}\left( \sqrt{\frac{r}{D}} L \right) f_{\rm odd}\left( \sqrt{\frac{r}{D}} \frac{L}{|a|} \right) + f_{\rm odd}\left( \sqrt{\frac{r}{D}} L \right) \left( r \kappa + f_{\rm even}\left(\sqrt{\frac{r}{D}} \frac{L}{|a|} \right) \right) \right] \;.
\end{equation}
As before, it is convenient to define the dimensionless MFPT $\tilde T(x) = D\,T(x)/L^2$, and in particular, the unknown boundary value
\bea \label{TtildeL}
\tilde{T}(-L/|a|)  = \frac{D\,\kappa}{L^2} \;.
\eea
The dimensionless MFPT $\tilde{T}(0) = D T(0)/L^2$ can then be expressed as a function of two dimensionless parameters $-1 < a \leq 0$ and $\beta = L \sqrt{r/D}$ and the unknown boundary value $\tilde{T}(-L/|a|)$ in Eq. (\ref{TtildeL}), leading to
\begin{equation} \label{eq:MFPT-neg-a-origin-dimensionless}
\tilde{T}(0) = \frac{1}{\beta^2}\frac{f_{\rm even}(\beta) f_{\rm odd}(\beta/|a|) + f_{\rm odd}(\beta) \left( \beta^2 \tilde{T}(-L/|a|) + f_{\rm even}(\beta/|a|) \right)}{\left[f_{\rm odd}\left(\beta/|a|\right) + f_{\rm odd}(\beta)\right]} \;.
\end{equation}

Thus the only unknown factor in the exact solution in Eq. (\ref{eq:MFPT-neg-a-origin-dimensionless}) is the single number $\tilde{T}(-L/|a|)  = D\,\kappa/L^2$. As explained in the beginning of this subsection, there is no simple way to determine this unknown boundary value $\kappa$ without solving for $T(x)$ on the full line, which is rather cumbersome unfortunately. Hence our strategy is to use the numerical value of $\tilde T(-L/|a|)$ obtained from simulations and then compare the analytical solution for $\tilde T(0)$ in Eq.~(\ref{eq:MFPT-neg-a-origin-dimensionless}) with the numerical answer for $\tilde T(0)$, for different values of the parameters $\beta$ and $a$. This is reported in Fig. \ref{Ttilde_aneg}, where we see a perfect agreement between the analytical $\tilde T(0)$ (with $\kappa$ as a numerical input) and the value of $\tilde T(0)$ obtained from simulations for different $\beta$ and $-1 < a \leq 0$.  
\begin{figure}[t]
\includegraphics[width = 0.9\textwidth]{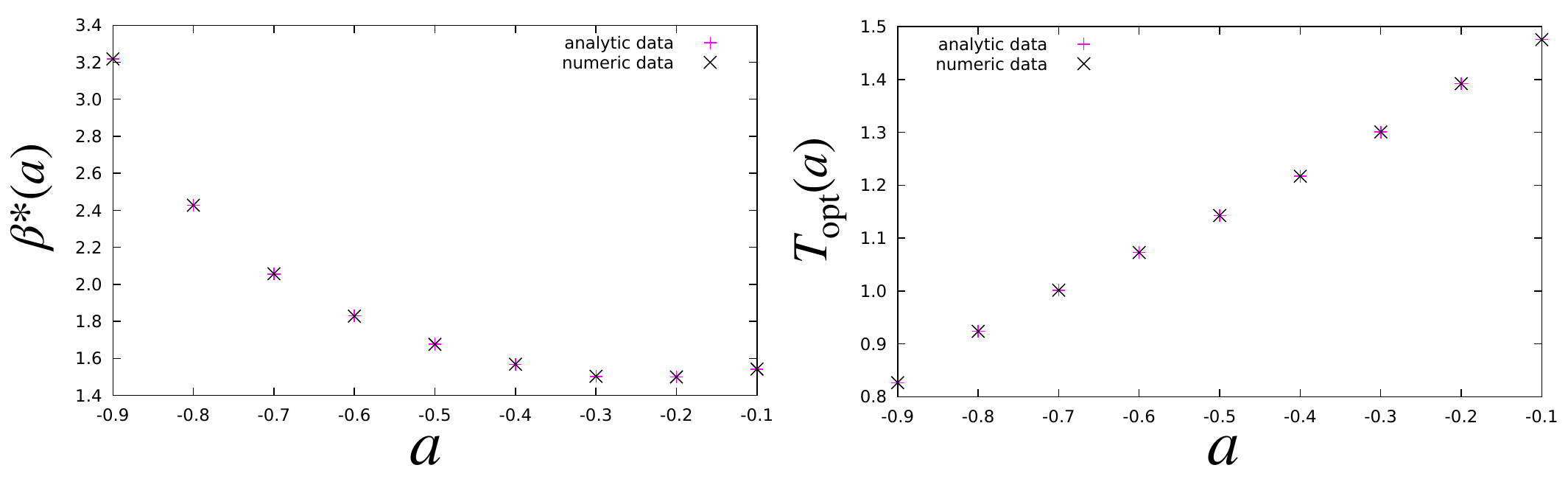}
\caption{{\bf Left:} The optimal value $\beta^*(a)$, at which $\tilde T(0)$ achieves its minimum as a function of $\beta$ for a fixed $a$, is shown for few values of $a$, for $-1 < a \leq 0$. {\bf Right:} The optimal MFPT $T_{\rm opt}(a)$, i.e., $\tilde T(0)$ evaluated at $\beta = \beta^*(a)$, for the same values of $-1 \leq a < 1$, as in the left panel. Unlike for $0 \leq a < 1$ in Fig. \ref{Figbetastarapos}, we do not show a continuous theoretical curve of $\beta^*(a)$ vs $a$ because, for $-1<a\leq 0$, for each value of $a$, we need to take $\kappa$ as an input from numerical simulations and this is done only for few select values of $a$. The right panel shows that the optimal MFPT $T_{\rm opt}(a)$, as a function of $a$, decreases compared to $a=0$. Thus a negative value of $a$ clearly reduces the optimal MFPT.}\label{Figbetastaraneg}
\end{figure}

Having fixed this only unknown $\kappa$ from the simulations, we have access to the full analytical formula for $\tilde T(0)$ as a function of $\beta$, for different values of $-1<a\leq 0$. As seen from Fig. \ref{Ttilde_aneg}, the MFPT $\tilde T(0)$ exhibits a minimum at $\beta = \beta^*(a)$. This optimal value $\beta^*(a)$ is plotted as a function of $-1 < a \leq 0$ in the left panel of Fig. \ref{Figbetastaraneg}. Finally, the optimised MFPT $T_{\rm opt}(a)$ (i.e., $\tilde T(0)$ evaluated at $\beta = \beta^*(a)$) is plotted as a function of $-1<a \leq 0$ in the right panel of Fig. \ref{Figbetastaraneg}. Contrary to the case $0\leq a < 1$, we see that, for $-1< a \leq 0$, the optimal MFPT decreases from its value at $a=0$ as $a$ decreases. This indicates that, a negative $-1<a\leq 0$ actually expedites the search process compared to the $a=0$ case, i.e., the standard resetting to the origin. 

As in the case of $0 \leq a < 1$, one can perform a small $a$ expansion of $\tilde T(0)$ in Eq. (\ref{eq:MFPT-neg-a-origin-dimensionless}) which, fortunately, does not require the knowledge of the unknown $\kappa$. Indeed, in the $a \to 0$ limit, the term $\beta^2 \tilde T(-L/|a|) \ll f_{\rm even}(\beta/|a|)$ on the right hand side of Eq. (\ref{eq:MFPT-neg-a-origin-dimensionless}). This is due to the fact that $f_{\rm even}(\beta/|a|) \sim e^{\beta/|a|}$ as $a \to 0$ -- this follows from Eq. (\ref{eq:def-feven-fodd}) since $C_{2n} \to {1}$ as $a \to 0$. In contrast, the term $\beta^2 \tilde T(-L/|a|)$ can not grow faster than $L^2/|a|^2$ as $a \to 0$ (this follows from the boundary condition at $x \to - \infty$ discussed earlier). Consequently, neglecting $\beta^2 \tilde T(-L/|a|)$ on the right hand side of Eq. (\ref{eq:MFPT-neg-a-origin-dimensionless}) and taking the $a \to 0$ limit in the functions $f_{\rm even}$ and $f_{\rm odd}$, we arrive at 
\begin{equation} \label{Tsmallaneg}
\tilde{T}(0) = \frac{1}{\beta^2} \left[e^\beta - 1 - |a| \beta\right] \mbox{~~when~~} |a| \ll 1 \;.
\end{equation}
In fact, as in the $0 \leq a < 1$ case, this result for small $|a|$ can also be derived directly from the differential equation (\ref{eq:MFPT-ODE}), as shown in Appendix \ref{App_smalla}. Indeed, the result in Eq. (\ref{Tsmallaneg}) is identical to Eq. (\ref{eq:MFPT-pos-a-a-asymptotics}) with $a$ replaced by $-|a|$ and hence this small $a$ expansion for the whole range $-1 < a < 1$ reads the same as (\ref{eq:MFPT-pos-a-a-asymptotics}). Consequently the analysis presented in (\ref{betastar_small_a}) and (\ref{Topt_small_a}) also holds for small negative $a$ as well.


\section{Numerical simulations}\label{sec:numerics}

To verify the analytical results for the MFPT discussed above, we use numerical simulations (for a pedagogical account see Ref.  \cite{practical_guide2015}).
For the sake of simplicity we set the diffusion constant to $D=1$. The naive way to simulate the rescaling random walker is by just 
numerically iterating Eq. (\ref{def_RBM}) with an appropriate time step $\dd t$.
To begin with, we apply a relatively large step
size of $\dd t=1$. For this purpose, we start the walk at position $x(t=0)=x_0$ (in our case either $x_0=0$ or $x_0=-L/|a|$).
Also, we draw the first duration $t_r$ until
rescaling  from the exponential distribution
with rate $r$. This means \cite{practical_guide2015},
we draw a random number $q$ uniformly distributed in $[0,1]$ and set
$t_r = - \log(1-q)/r$. Then we simulate our walker by always using the step
size of $dt$, except for the case
the rescaling event will happen after the current time $t$
but before the next
considered time $t+dt$. In this case, the time step is $t+dt-t_r$.
For a scaling event,
we multiply the position $x$ with $a$ after said motion step was
performed and draw a new duration until rescaling.

This process is continued until
the target, located at $x=L$, is reached. This means, for the given time resolution,
the target is crossed between two consecutive positions. Thus, it is tested
whether the walker changes its relative position with respect to $x=L$,
i.e.,  $(L-x(t))(L-x(t+\dd t))<0$ holds, but only for the case no rescaling
happens between the times $t$ and $t+dt$.
The first-passage time (FPT) obtained this way could be averaged over independent runs.
However, this will systematically overestimate the MFPT, since for any two
consecutive points of the sequence,
i.e., $[x(t), x(t+\dd t)]$, there is a finite chance that the walker reached
the target in between, even 
if both points are below or both are above $L$.
In fact, this probability is given by (see Appendix \ref{app:crossing})
\begin{equation}
    \label{probability_equation}
    p_{\rm pass}(x(t+\dd t), x(t), \dd t, L) = \min\left(1, \exp \left(\frac{-(L-x(t))(L-x(t+\dd t))}{\dd t}\right)\right)\,.
\end{equation}

This systematic error can be reduced by using smaller step sizes $\dd t$, but this 
quickly becomes unfeasible due to the increase in required computational
effort.
Instead we can apply an approach which is based on iteratively refining
parts of the walk which might involve a first passage \cite{Walter2020}.
The main idea is to successively refine intervals $[x(t), x(t + \dd t)]$ by sampling an intermediary point $x\left(t+\frac{\dd t}{2}\right)$ with the correct statistics.
This can be done via so-called \emph{generalized
  Brownian bridges}, which are, for the discrete-time case, random walks
$\hat x(t)$ between two specified points, initially $\hat x(t_i)=x_i$ and
finally $\hat x(t_f)=x_f$. Such a Brownian bridge with {\it discrete} steps can be obtained
by first generating a standard
random walk $\tilde x(t)$ with initial condition $\tilde x(t_i)=0$ and then
setting \cite{deBruyne2021}
\begin{equation}
  \hat x(t)= x_i + \tilde x(t) -
  \frac{t-t_i}{t_f-t_i}(\tilde x(t_f)-(x_f-x_i))\quad (t_i\le t \le t_f)\,.
  \label{eq:brownian:bridge}
\end{equation}
Clearly, $\hat x(t_i)=x_i$ and $\hat x(t_f)=x_f$ as required, and one can show
that $\hat x(t)$ has the correct statistics. Here, we are interested in
a Brownian bridge for the case $t_i\equiv t$, $t_f\equiv t+\dd t$,
$x_i\equiv x(t)$, $x_f\equiv x(t+\dd t)$
and obtaining just the position for time $t+\frac 1 2 \dd t$. Thus, to generate
the desired bridge, we need an auxiliary two-step random walk $\hat x(t)$ with
step size $\frac 1 2 \dd t$, which we write as
\begin{align}
    \hat x\left( \dd t/2\right )  &= \eta_1 \sqrt{\dd t} \equiv m_1 \label{eq_m1}\\
    \hat x\left( \dd t \right) &= m_1 + \eta_2 \sqrt{\dd t} \equiv m_2 \label{eq_m2}\,,
\end{align}
where  $\eta_1$ and $\eta_2$ are independent 
centered unit-variance Gaussian random variables.
By inserting Eqs. (\ref{eq_m1}) and (\ref{eq_m2}) into Eq. (\ref{eq:brownian:bridge}) one obtains
\begin{equation}
    \label{bisection_eq}
    x\left(t+\frac{\dd t}{2}\right) = x(t) + m_1 - \frac{1}{2}
    \left(m_2-\left(x(t+\dd t)-x(t)\right)\right).
\end{equation}

The main idea of the algorithm is now to first sample
the rescaling random walker 
with a rather large step size, e.g., $\dd t=1$, and then refine the
obtained walk by applying 
\eqref{bisection_eq} in intervals where the probability that the target 
was reached in between, given by \eqref{probability_equation}, is 
larger than some threshold $\theta \ll 1$. Here we typically
used a value of $\theta = 10^{-10}$, only for few test cases
smaller values down to $\theta = 10^{-30}$. In Appendix \ref{sec:algorithms}, we present the approach with a higher
level of detail by stating two algorithms.
The first algorithm is used to generate the initial walk and
setting up a queue with tentative time-and-space intervals which are possibly refined.
The
second algorithm then iteratively takes intervals from the queue,
refines them if needed, resulting in new smaller intervals,
and removes from the queue intervals which are beyond the so-far
detected FPT.

\section{Conclusion} \label{sec:conclusion}

In this paper, we have studied a simple model of a diffusive particle on a line, undergoing a stochastic resetting with rate $r$, 
via rescaling its current position by a factor $a$. For the case $|a|<1$, the system approaches a nonequilibrium stationary state at long times
where the position distribution becomes stationary. We have computed this position distribution
analytically and verified it numerically for all $|a|<1$. The stationary position distribution
in this case is symmetric with an exponential tail for large argument $|x|$, while having a Gaussian shape near its peak at $x=0$. We also studied the
mean first-passage time (MFPT) $T(0)$ to a target located at a distance $L$ from the initial position (the origin) of the particle. While it is easy to write a backward differential equation for $T(x)$ (denoting the MFPT starting from the initial position $x$), this equation is hard to solve since it is nonlocal in space. Nevertheless, for $0\leq a < 1$, we managed to obtain an exact analytical expression for $T(0)$ and showed that, while it has a minimum at an optimal value $r^*$, the corresponding optimal MFPT increases with $a$ for
$a \in [0,1)$, indicating that a positive rescaling ($a>0$) is not beneficial for the search of the target, compared to the standard resetting to the origin ($a=0$). 
In contrast, for $a \in (-1,0]$, we have shown that the optimal MFPT decreases as $a$ decreases compared to $a=0$, demonstrating that rescaling by a negative factor, i.e., a rescaling followed by a reflection around the origin, is a better search strategy than the standard resetting to the origin.     

There are several interesting directions in which one can extend this work. For instance, it would be interesting to study the position distribution
and the MFPT for this rescaling-diffusive process in two or higher dimensions. Another interesting question is what happens to the position distribution when one switches on an external confining potential $U(x)$? Furthermore, this rescaling-resetting process may also be studied for non-diffusive stochastic processes, such as L\'evy walks \cite{BCHPM23} and for active run-and-tumble particles~\cite{Berg04,TC08,BLLRVV16}.

\acknowledgements
MB, SNM and GS acknowledge support from ANR, Grant No. ANR- 23- CE30-0020-01 EDIPS. SNM acknowledges the Alexander von Humboldt foundation for the Gay Lussac-Humboldt prize that allowed a visit to the Physics department at Oldenburg University, Germany where this work initiated. We thank K. S. Olsen for pointing out that the MFPT in this problem, but only in the range $0\le a<1$, was also studied in unpublished works
by K. S. Olsen and H. L\"owen and independently by C. Di Bello, R. Metzler
and Z. Palmowski.
\appendix

\section{Expansion for small $a$}\label{App_smalla}

We start from Eq. (\ref{eq:MFPT-ODE}). For small $a$, we expand the nonlocal term $T(a\,x)$ up to linear order in $a$, giving
\bea \label{small_a.1}
T(a\,x) \approx T(0) + a\,x \,T'(0) + O(a^2) \;,
\eea
where $T(0)$ and $T'(0)$ are yet to be determined. Substituting this expansion in Eq. (\ref{eq:MFPT-ODE}) reduces the nonlocal equation to an ordinary local second order differential equation but with inhomogeneous unknown terms that need to be determined self-consistently. This equation reads
\bea \label{small_a.2}
D\,T''(x) - r\,T(x) = - 1 - r\,T(0) - a\,r\,T'(0)\, x \;.
\eea
This equation needs to be solved in the region $x \neq L$, with the boundary conditions $T(L)= 0$ and that $T(x)$ does not grow faster than $\sim x^2$ as $x \to - \infty$. Performing the shift 
\bea \label{shift}
T(x) = \frac{1+r\,T(0)+a\,r\,T'(0)\,x}{r} + W(x) \;,
\eea
gives a homogenous differential equation for $W(x)$
\bea \label{eq_W}
D\, W''(x) - r\, W(x) = 0 \;.
\eea
The only acceptable solution that does not grow faster than $\sim x^2$ as $x \to - \infty$ is given by $W(x) = A\, e^{\sqrt{r/D}\, x}$ where $A$ is yet to be determined. Thus one has the full solution
\bea \label{W_full}
T(x) =  \frac{1+r\,T(0)+a\,r\,T'(0)\,x}{r} + A\, e^{\sqrt{r/D}\, x} \;.
\eea 
The boundary condition $T(L) = 0$ fixes the constant $A$ and we get
\bea \label{W_full.2}
T(x) =  \frac{1+r\,T(0)+a\,r\,T'(0)\,x}{r}  -  \frac{1+r\,T(0)+a\,r\,T'(0)\,L}{r} e^{\sqrt{r/D}\, (x-L)} \;.
\eea
Taking a derivative with respect to $x$ gives 
\bea \label{Tprime}
T'(x) = a\, T'(0) - \sqrt{\frac{r}{D}}\, \frac{1+r\,T(0)+a\,r\,T'(0)\,L}{r} e^{\sqrt{r/D}\,(x-L)} \;.
\eea
Setting $x=0$ in Eqs. (\ref{W_full.2}) and (\ref{Tprime}) gives two self-consistent linear equations for the
two unknowns $T(0)$ and $T'(0)$. Solving this, trivially, we get, for the dimensionless MFPT defined as $\tilde T(0) = D T(0)/L^2$
\bea \label{T_final}
\tilde{T}(0) \simeq \frac{1}{\beta^2} \left[e^\beta - 1  + a \beta + O(a^2)\right] \quad, \quad {\rm as} \quad a \to 0 \;,
\eea
which clearly holds for both positive and negative $a$.

\section{Crossing probability}\label{app:crossing}

\begin{figure}[t]
\includegraphics[width = 0.5\linewidth]{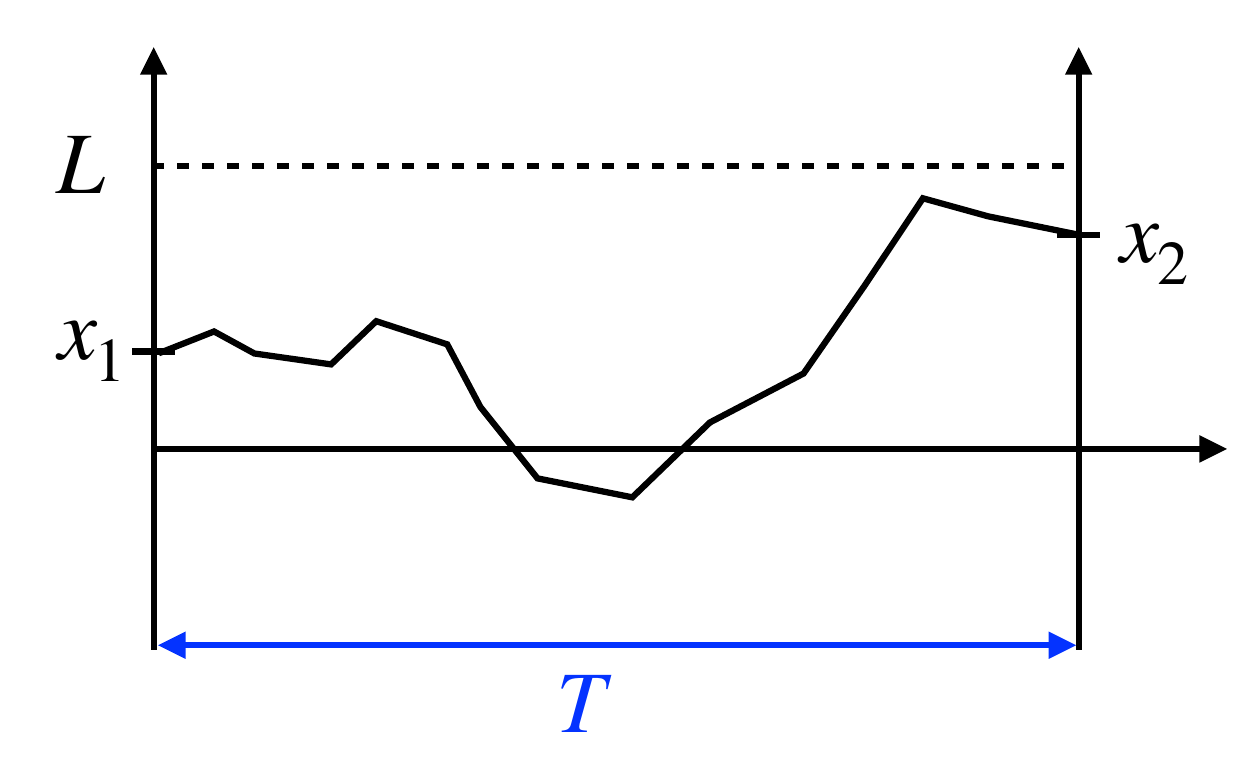}
\caption{A schematic trajectory of a Brownian motion propagating from $x_1$ to $x_2$ in time $T$. We calculate below the conditional probability that the walker passes through the level $L$ given that it has reached $x_2$ at time $T$.}\label{Fig_crosssing}
\end{figure}

We consider a one-dimensional Brownian motion propagating from $x_1$ to $x_2$ in time $T$. The Brownian propagator, i.e., the probability density to reach $x_2$ at time $T$, starting from $x_1$, is given by
\bea \label{BM_propag}
G_0(x_2,T|x_1,0) = \frac{1}{\sqrt{4 \pi D\,T}}\exp{\left(-\frac{(x_2-x_1)^2}{4D\,T}\right)} \;.
\eea 
We want to calculate the probability that the walker passes through $x=L$, given that it has reached $x_2$ at time $T$. We assume $x_1 \leq L$. Clearly, if $x_2 \geq L$, this conditional probability is just one, since the particle has to cross the level $x=L$ in order to reach $x_2 \geq  L$. Now consider the complementary case where $x_2 < L$. To compute the passing probability in this case, it is convenient instead to compute the conditional survival probability, i.e., the probability $S(x_2, T|x_1,0)$ that the particle does not cross (survive) the level $L$, given that it reached $x_2 < L$ at time $T$. For this purpose,
we consider  
unconditioned survival probability $S_0(x_2, T|x_1,0)$, i.e. that the walker
survives and reaches position $x_2$ at time $T$.  It can be computed by solving the diffusing equation for $x < L$ with an absorbing boundary condition at $x=L$ \cite{Redner_book,BMS13}. The solution can be easily obtained by the method of images \cite{Redner_book,BMS13} and it reads 
\bea \label{survival}
S_0(x_2, T|x_1,0) = \frac{1}{\sqrt{4 \pi D T}} \left( e^{-{(x_2-x_1)^2}/{(4DT)}} - e^{-{(2L-x_2-x_1)^2}/{(4DT)}}\right) \;.
\eea 
The conditional survival probability is then given by the ratio
\bea \label{survival_cond.1}
S(x_2, T|x_1,0) = \frac{S_0(x_2, T|x_1,0)}{G_0(x_2, T|x_1, 0)} = 1 - e^{-(L-x_1)(L-x_2)/(D\,T)} \;.
\eea
Hence the passing probability $p_{\rm pass}(x_2, x_1, T,L)$ is given by the complement
\bea \label{P_pass}
p_{\rm pass}(x_2, x_1, T,L) = 1 - S(x_2, T|x_1,0) = e^{-(L-x_1)(L-x_2)/(D\,T)} \quad, \quad {\rm for} \quad x_2 < L \;.
\eea
Hence, considering both cases $x_2 \geq L$ and $x_2 < L$, the passing probability can be written in a compact form
\bea \label{P_pass_compact}
p_{\rm pass}(x_2, x_1, T,L) = \min \left(1, \exp\left(-\frac{(L-x_1)(L-x_2)}{D\,T}\right)\right) \;.
\eea
Setting $x_2 = x(t+\dd t)$, $x_1=x(t)$, $T=\dd t$ and $D=1$ gives Eq. (\ref{probability_equation}) in the main text.  

\section{Detailed algorithm \label{sec:algorithms}}

The algorithms we applied work in detail as follows.

The algorithms maintain a queue of time-and-space intervals. Only those intervals
are stored
which with high-enough probability contain
a crossing of the target $x=L$.
The elements of the queue are stored as $[x(t), x(t+\dd t), t, \dd t,$ depth$]$,
ordered by increasing times $t$. The time span of the interval for the initial
queue is usually given by the global value $\dd t$, here $\dd t=1$. But the
initial queue
will also contain some intervals with smaller time spans which appear
just before the rescaling events.
Furthermore, for the queue of initial intervals, the depth is always 0. During
bisection, i.e., whenever an interval
is split into two sub intervals later on, the depth is increased by one.

The algorithms have two global parameters. First, $\theta$ states the probability threshold of
missing a crossing of the target, actually at least a double crossing.
We also use a parameter  $\omega$ which is
the maximum number of splits allowed on any of the initial intervals,
i.e., the maximum depth.
This is needed because for the interval which contains our target $L$,
\eqref{probability_equation} will yield 1 
and thus the algorithm would never finish without the criterion based
on the number of splits.

The pseudo code of the algorithms can be found below. First, the initial queue is
calculated via 
\autoref{alg1}. Its main loop, which runs until the target at $x=L$ is found
at the given time resolution $\dd t$, contains of two parts.
In the first part, steps
of size $\dd t$ are performed and intervals containing a passage with high
enough probability are added to the queue. This part runs
until a rescaling occurs.
Next, the second part of the loop advances
the walk one step until the actual rescaling time, i.e. for a smaller time than $\dd t$.
Next, the  rescaling is performed and the main loop
continues.

The interval in which the estimated first 
passage occurs, depending on the threshold probability  $\theta$,
is obtained by  \autoref{alg2}. Here, iteratively intervals
are taken from the queue and split. Note that whenever an interval contains
a crossing of the target $x=L$, all intervals with higher time in the queue can be removed.
The two halves are treated. If the first half contains $x=L$, this half will
be put in the queue for later consideration, if the depth is not too large,
the second half can be discarded.
If only the second half contains $x=L$, it will be put into the queue.
Technically, this happens
in the last part of the loop by checking $p_\text{cross}$, because
$p_\text{cross}=1$ holds in this case.
Here, also the first halve of the interval is added to the queue if the
probability $p_\text{cross}>\theta$, because
it still might contain the first passage with high enough probability. If for none
of the two halves a crossing of $x=L$ is detected so far,
each half will be further considered only
if $p_\text{cross}$ is large enough, respectively.

After \autoref{alg2} is finished,
the actual FPT can now either be interpolated from the interval
containing the passage event, or one can just take the stored time 
value directly, since $\dd t$ will usually be very very small. This
is the case at least for the
parameters $\omega=100$ and $\theta=10^{-10}$ we usually applied.

To calculate the MFPT we can now average over an appropriate number
of samples.

\begin{algorithm}[h]
    \setstretch{1}
    \CommentSty{\color{blue}}
    \caption{Initialize queue for bisection}\label{alg1}
             \SetKwInOut{Input}{input}
            \SetKwInOut{Output}{output}
            \Input{$x_0$, $r$, $a$, $L$, $\theta$}
            \Output{queue, fpt\_interval\_estimate}
        \SetKwBlock{Beginn}{beginn}{ende}
        \Begin{
            $t=0$\;
            $dt=1$\tcp*{you can use other step sizes here}
            $t_r$ = draw random number from exponential distribution with rate $r$\;
            current\_x = $x_0$\;
            \textbf{Initialize} queue\tcp*{Creates a variable for a yet empty queue}
            \textbf{Initialize} fpt\_interval\_estimate\tcp*{stores interval of the first passage}
            \While{true}{
                \While{$t+d t< t_r$\tcp*{advance walk without rescaling}}{
                    $\eta$ = draw zero-mean unit-variance Gaussian random variable\;
                    next\_x = current\_x + $\eta \sqrt{2\,dt}$\;
                    \If{$p_{\rm cross}$\emph{(current\_x, next\_x, dt, L)} $> \theta$
                    \tcp*{see \autoref{probability_equation}}}{\textbf{put} [current\_x, next\_x, t, dt, 0] at the end of queue\;
                        \If{\emph{[current\_x, next\_x] contains L}}{
                            fpt\_interval\_estimate = [current\_x, next\_x, t, dt]\;
                            \Return queue, fpt\_interval\_estimate
                        }
                    }
                    t += dt\;
                    current\_x = next\_x\;
                }
                rest\_time = $t_r - t$\tcp*{advance walk to next rescaling}
                $\eta$ = draw zero-mean unit-variance Gaussian random variable\;
                next\_x = current\_x + $\eta \sqrt{2\, \text{rest\_time}}$\;
                \If{$p_{\rm cross}$\emph{(current\_x, next\_x, dt, L)} $> \theta$\tcp*{see \autoref{probability_equation}}}{
                    \textbf{put} [current\_x, next\_x, t, rest\_time, 0] at the end of queue\;
                    \If{\emph{[current\_x, next\_x] contains L}}{
                        fpt\_interval\_estimate = [current\_x, next\_x, t, rest\_time]\;
                        \Return queue, fpt\_interval\_estimate
                    }
                }
                $t$ += rest\_time\;
                current\_x = $a * \text{next\_x}$\;
                $t_r$ = draw random number from exponential distribution with rate $r$\;
                $t_r$ += $t$\;
            }
        }
\end{algorithm}

\newpage

\begin{algorithm}[h]
    \setstretch{1}
    \CommentSty{\color{blue}}
    \caption{Refine first passage time by bisections}\label{alg2}
             \SetKwInOut{Input}{input}
            \SetKwInOut{Output}{output}
            \Input{$L$, $\theta$, $\omega$, queue, fpt\_interval\_estimate}
            \Output{fpt\_interval}
        \SetKwBlock{Beginn}{beginn}{ende}
        \Begin{
            fpt\_interval = fpt\_interval\_estimate\;
            \While{\emph{queue not empty}}{
                [$x_1, x_2, t$, dt, depth] = queue.pop()\tcp*{take front element from queue}
                $\eta_1$ = draw zero-mean unit-variance Gaussian random variable\;
                $\eta_2$ = draw zero-mean unit-variance Gaussian random variable\;
                $m_1=\eta_1\sqrt{\text{dt}}$\;
                $m_2=m_1 + \eta_2\sqrt{\text{dt}}$\;
                mid = $x_1+m_1-0.5 (m_2-(x_2-x_1))$\;
                \If{\emph{[$x_1$, mid] contains L}}{
                    \textbf{clear} queue\tcp*{all later intervals discarded}
                    fpt\_interval = [$x_1$, mid, $t$, dt/2]\;
                    \If{\emph{depth + 1} $<\omega$} {
                        \textbf{put} [$x_1$, mid, $t$, dt/2, depth+1] \textbf{in} queue
                    }
                    \textbf{continue}\tcp*{jump to beginning of while loop}
                }
                
                \If{\emph{[mid, $x_2$] contains L}}{
                    \textbf{clear} queue\tcp*{all later intervals discarded}
                    fpt\_interval = [mid, $x_2$, $t+dt/2$, dt/2]\tcp*{is put in queue below}
                }
                \If{\emph{depth + 1} $< \omega$}{
                    \tcc{Note: If $L\in$ [mid, $x_2$] then p(mid, $x_2$, dt/2, L)=1}
                    \If{$p_\text{cross}$\emph{(mid, $x_2$, dt/2, L)}$> \theta$\tcp*{see \autoref{probability_equation}}}{
                        \textbf{put} [mid, $x_2$, t+dt/2, dt/2, depth+1] at front of queue\;
                    }
                    \If{$p_\text{cross}$\emph{($x_1$, mid, dt/2, L)}$> \theta$}{
                        \textbf{put} [$x_1$, mid, t, dt/2, depth+1] at front of queue\;
                    }
                }
            }
            \Return fpt\_interval
        }
    \end{algorithm}


\clearpage


\begin{thebibliography}{26}

\bibitem{VLRS11}
G. M. Viswanathan, M. G. da Luz, E. P. Raposo, and H. E. Stanley, {\it The physics of foraging: an introduction to random searches and biological encounters} (Cambridge University Press, 2011)

\bibitem{BLMV11}
O. B\'enichou, C. Loverdo, M. Moreau, R. Voituriez, Rev. Mod. Phys. {\bf 83}, 81 (2011)

\bibitem{EM11} M. R. Evans, S. N. Majumdar, Phys. Rev. Lett. {\bf 106}, 160601 (2011)
\bibitem{EM11b} M. R. Evans, S. N. Majumdar, J. Phys. A: Math. Theor. {\bf 44}, 435001 (2011)
\bibitem{TPSRR20} O. Tal-Friedman, A. Pal, A. Sekhon, S. Reuveni, Y. Roichman, J. Phys. Chem. Lett. {\bf 11},7350 (2020)
\bibitem{EMS20} M. R. Evans, S. N. Majumdar, G. Schehr, J. Phys. A: Math. Theor. 53, 193001 (2020)
\bibitem{GJ2022}
S. Gupta, A. M. Jayannavar, Frontiers in Physics {\bf 10}, 789097 (2022)
\bibitem{PKR2022}
A. Pal, S. Kostinski, S. Reuveni, J. Phys. A: Math. Theor. {\bf 55}, 021001 (2022)
\bibitem{BBPMC20} B. Besga, A. Bovon, A. Petrosyan, S. N. Majumdar, S. Ciliberto, Phys. Rev. Res. {\bf 2},032029 (2020).

\bibitem{FBPCM21} F. Faisant, B. Besga, A. Petrosyan, S. Ciliberto, S. N. Majumdar, J. Stat. Mech.: Theory Exp., 113203 (2021).
\bibitem{TRR22} O. Tal-Friedman, Y. Roichman, S. Reuveni, Phys. Rev. E {\bf 106}, 054116 (2022)
\bibitem{P22} J. K. Pierce, arXiv preprint arXiv:2204.07215 (2022)
\bibitem{BCHPM23} C. Di Bello, A. V. Chechkin, A. K. Hartmann, Z. Palmowski and R. Metzler, N. J. Phys. {\bf 25}, 082002 (2023)
\bibitem{OG24}
K. S. Olsen, D. Gupta, J. Phys. A: Math. Theor. {\bf 57}, 245001 (2024)
\bibitem{H23} U. Harbola, Phys. Rev. E {\bf 108}, 014135 (2023)
\bibitem{HW89}
A. J. Hall, G. C. Wake,  J. Austral. Math. Soc. Ser. B {\bf 30}, 424 (1989)

\bibitem{DM06}
D. S. Dean, S. N. Majumdar, J. Stat. Phys. {\bf 124}, 1351 (2006)


\bibitem{Lar04}
H. Larralde, J. Phys. A {\bf 37}, 3759 (2004)
\bibitem{MK07}
S. N. Majumdar, M. J. Kearney, Phys. Rev. E {\bf 76}, 031130 (2007)

\bibitem{CSMS17}
A. V. Chechkin, F. Seno, R. Metzler, I. M. Sokolov, Phys. Rev. X {\bf 7}, 021002 (2017)

\bibitem{LG18}
Y. Lanoisel\'ee, D. S. Grebenkov, J. Phys. A: Math.Theor. {\bf 51}, 145602 (2018)

\bibitem{BB20}
E. Barkai, S. Burov, Phys. Rev. Lett. {\bf 124}, 060603 (2020)

\bibitem{GMS23}
M. Gu\'eneau, S. N. Majumdar, G. Schehr,  J. Phys. A: Math. Theor. {\bf 56}, 475002, (2023)

\bibitem{WABG09}
B. Wang, S. M. Anthony, S. C. Bae, S. Granick, Proc. Natl. Acad. Sci. USA {\bf 106}, 15160 (2009)


\bibitem{WGLFGH19}
P. Witzel, M. G\"otz, Y. Lanoisel\'ee, T. Franosch, D. S. Grebenkov, D. Heinrich, Biophys. J. {\bf 117}, 203 (2019)


\bibitem{practical_guide2015}
A. K. Hartmann, {\it Big Practical Guide to Computer Simulations}, (World Scientific, Singapore, 2015)


\bibitem{Walter2020}
B. Walter, K. J. Wiese, Phys. Rev. E {\bf 101}, 043312 (2020)


\bibitem{deBruyne2021}
B. De Bruyne, S. N. Majumdar, G. Schehr, Phys. Rev. E {\bf 104}, 024117 (2021)


\bibitem{Redner_book}
S. Redner, {\it A Guide to First-Passage Processes} (Cambridge University Press, Cambridge, UK, 2007)

\bibitem{BMS13}
A. J. Bray, S. N. Majumdar, G. Schehr, Adv. Phys. {\bf 62}, 225 (2013)




\bibitem{Berg04}
H. C. Berg, {\it E. Coli in Motion}, (Springer Verlag, Heidelberg, Germany) (2004)

\bibitem{TC08}
J. Tailleur, M. E. Cates, Phys. Rev. Lett. {\bf 100}, 218103 (2008)

\bibitem{BLLRVV16}
C. Bechinger, R. Di Leonardo, H. L\"owen, C. Reichhardt, G. Volpe, G. Volpe, Rev. Mod. Phys. {\bf 88}, 045006 (2016)


%
%
%
%
%
%
%
%
%
%
%
%








\end{thebibliography}
\end{document}